\documentclass[preprint,showpacs,preprintnumbers,amsmath,amssymb]{revtex4}
\usepackage{booktabs}
\usepackage{mathrsfs}
\usepackage{epsfig}
\usepackage{graphicx}
\usepackage{dcolumn}
\usepackage{bm}
\usepackage{amsmath}
\usepackage{slashed}       
\usepackage{amssymb}
\usepackage{amsfonts}

\let\jnfont=\rm
\def\NPB#1,{{\jnfont Nucl.\ Phys.\ B }{\bf #1},}
\def\PLB#1,{{\jnfont Phys.\ Lett.\ B }{\bf #1},}
\def\EPJC#1,{{\jnfont Eur.\ Phys.\ Jour.\ C }{\bf #1},}
\def\PRD#1,{{\jnfont Phys.\ Rev.\ D }{\bf #1},}
\def\PRL#1,{{\jnfont Phys.\ Rev.\ Lett.\ }{\bf #1},}
\def\MPLA#1,{{\jnfont Mod.\ Phys.\ Lett.\ A }{\bf #1},}
\def\JPG#1,{{\jnfont J.\ Phys.\ G}{\bf #1},}
\def\CTP#1,{{\jnfont Commun.\ Theor.\ Phys.\ }{\bf #1},}
\def\ZPC#1,{{\jnfont Z.\ Phys.\ C }{\bf #1},}
\def\JHEP#1,{{\jnfont JHEP \ }{\bf #1},}
\def\lsim{\raise0.3ex\hbox{$<$\kern-0.75em\raise-1.1ex\hbox{$\sim$}}}
\def\gsim{\raise0.3ex\hbox{$>$\kern-0.75em\raise-1.1ex\hbox{$\sim$}}}

\newcommand{\2}{\ \ }
\newcommand{\8}{\ \ \ \ \ \ \ \ }

\begin{document}
\preprint{\parbox{1.2in}{\noindent arXiv:1303.2426}}

\title{The SM extension with color-octet scalars:
diphoton enhancement and global fit of LHC Higgs data}

\author{Junjie Cao$^{1,2}$, Peihua Wan$^1$, Jin Min Yang$^3$, Jingya Zhu$^3$}

\affiliation{
  $^1$  Department of Physics,
        Henan Normal University, Xinxiang 453007, China \\
  $^2$ Center for High Energy Physics, Peking University,
       Beijing 100871, China \\
  $^3$ State Key Laboratory of Theoretical Physics,
      Institute of Theoretical Physics, Academia Sinica, Beijing 100190, China
      \vspace{1cm}}

\begin{abstract}
In light of the significant progress of the LHC to determine the properties
of the Higgs boson, we investigate the capability of the Manohar-Wise model
in explaining the Higgs data. This model extends the SM by one family of
color-octet and isospin-doublet scalars, and it can sizably alter the coupling
strengths of the Higgs boson with gluons and photons.
We first examine the current constraints on the model,
which are from unitarity, the LHC searches for the scalars and the electroweak precision data (EWPD).
In implementing the unitarity constraint, we use the properties of the SU(3) group to simplify the calculation.
Then in the allowed parameter space we perform a fit of the model, using
the latest ATLAS and CMS data, respectively. We find that the Manohar-Wise model is able to
explain the data with $\chi^2$ significantly smaller than the SM value. We also find that 
the current Higgs data, especially the ATLAS data,
are very powerful in further constraining
the parameter space of the model. 
In particular, in order to explain the $\gamma \gamma$ enhancement reported by the ATLAS
collaboration, the sign of the $hgg$ coupling is usually opposite to that in the SM.
\end{abstract}


\maketitle

\section{Introduction}
Based on about 25 fb$^{-1}$ data collected at 7-TeV and 8-TeV LHC, the ATLAS and
CMS collaborations have further confirmed the existence of a new boson with the
local statistical significance reaching $9 \sigma$ and more than $7\sigma$ respectively
\cite{1207-a-com,1207-c-com,1303-atlas-Moriond,1303-c-com}.
So far the mass of the boson is rather precisely determined to be around 125 GeV,
and its other properties, albeit with large experimental uncertainties,
agree with these of the Higgs boson in the Standard Model
\cite{1303-c-com,1303-a-com}. Since such a Higgs-like boson should play a role in the electroweak symmetry breaking and the mass
generation, its discovery is widely considered as a great success of the LHC and
marks a milestone in understanding the mechanism of the electroweak symmetry breaking.
But on the other hand, such
a discovery also poses some new questions. For example, as the experimental precision
to determine the properties of this Higgs-like boson has been improved significantly,
it is urgent for theorists to investigate the nature of this boson,
especially its role in the electroweak symmetry breaking and mass generation.
To answer these questions, various methods
have been proposed to extract physical information of this boson from the LHC data
\cite{hfit-Azatov,hfit-egmt,hfit-gkrs,hfit-ey,hfit-Corbett,hfit-os,HiggsSignal},
which showed that the current data, especially the sizable deviation of the $\gamma\gamma$
signal from its SM prediction \cite{1303-a-2ph,1303-c-2ph}, prefer new physics interpretation.
This conclusion makes it important to
explore the properties of the Higgs boson in various new physics models.

As the simplest modification of the SM Higgs sector, the two-Higgs-doublet model
has been extensively studied for almost thirty years. In its traditional realization
(called T2HDM hereafter), the model extends the SM by one family of color-singlet
and weak-doublet scalars. As a result, the model respects the custodial symmetry
in a natural way, avoids tree level flavor changing neutral current (FCNC) by
imposing a discrete $Z_2$ symmetry, and has interesting phenomenology at colliders
due to its rich spectrum in the scalar sector.
Because of these attractive features, the analysis of the Higgs
data in the T2HDM have been carried out since the first hint of the Higgs boson
at the LHC was released at the end of 2011
\cite{2hdm-Higgs,2hdm-extend,2hdm-lightHC,2hdm-aligned,2hdm-Gunion,1212-Gunion,
2hdm-Ferreira,hfit-1303-Giardino}. These studies, however, indicate that the T2HDM is not much better than the SM
in explaining the data (the extensions with new particles \cite{2hdm-extend}
or the aligned T2HDM \cite{2hdm-aligned} may be exceptions).
For example, in its most popular type-I and type-II versions, it has been shown that,
after considering various experimental and theoretical constraints, the T2HDM can explain at $1 \sigma$ level
the LHC data  only in a very narrow parameter space \cite{2hdm-Ferreira}, and the global minimum of
$\chi^2$ is roughly equal to the SM value \cite{hfit-1303-Giardino}. Confronted with
such a situation, we in this work investigate the prospect to explain the Higgs data
in another type of two-Higgs-doublet model, which is usually called Manohar-Wise
model \cite{cos-Model}. This model, well motivated by the principle of minimal
flavor violation, extends the SM by one family of scalars in the $(8,2)_{1/2}$
representation under the SM gauge groups. It retains the virtues of the T2HDM,
but it may explain the data in a more flexible way. To be more specific, in the T2HDM the only
way to influence the Higgs signal rates at the LHC is through modifying the decay rates of the Higgs boson
\cite{2hdm-Higgs,2hdm-lightHC,2hdm-aligned,2hdm-Gunion,1212-Gunion,2hdm-Ferreira}.
while the Manohar-Wise model can also alter the Higgs production rate at the LHC by changing the Higgs coupling
with gluons. Although this feature has been noticed before \cite{h-cos-pro,h-cos-xx,h-cos1,h-cos2,h-cos3,hh-cos},
the systematic study of Higgs properties in the Manohar-Wise model has not been performed.

It should be emphasized that the color-octet scalars are well motivated in many
basic theories, such as various SUSY constructions \cite{cos-susy}, topcolor
models \cite{cos-tc} and the models with extra dimensions \cite{cos-edm}.
Meanwhile, their phenomenology has been studied comprehensively. For example, the
single and pair productions of these scalars at the LHC were studied in \cite{cos-jj, cos-4j},
their implications in Higgs phenomenology were investigated in \cite{h-cos-pro,h-cos-xx,h-cos1,h-cos2,h-cos3,hh-cos},
and they were also utilized to explain the '$Wjj$' anomaly observed by the CDF \cite{cos-Wjj}.
In this work, we intend to investigate the capability
of the Manohar-Wise model to explain the Higgs data.
For this end, we first examine the theoretical and experimental
constraints on the model, which are from unitarity, the LHC searches for these scalars
and the electroweak data. Then we perform a fit to the current Higgs data. In implementing
the unitarity constraint on the model, we use the properties of the $SU(3)$ group to simplify
the calculation. This method, within our knowledge, has not been considered before.

The outline of the paper is as follows. In Section II, we briefly review the Manohar-Wise
model, and in Section III we discuss the unitarity and collider constraints on the model.
A fit of the model to the current Higgs data is performed in Section IV and the behavior of
the model to explain the data is illustrated. Finally, we present our conclusion in Section V.

\section{The Manohar-Wise Model}
Motivated by the principle of minimal flavor violation,
the Manohar-Wise model extends the SM by one family
of color-octet scalars in the $(8,2)_{1/2}$ representation of the gauge group
$SU(3)_C \bigotimes SU(2)_L \bigotimes U(1)_Y$ \cite{cos-Model}.
The explicit form of the scalars is given by
\begin{eqnarray}
S^{A}=\left( \begin{array}{c}
        S^A_+ \\
        \frac{1}{\sqrt{2}} (S^A_R+ i S^A_I)
      \end{array} \right),
\end{eqnarray}
where $A=1,...,8$ is color index, $S^A_+$ denotes a electric charged color-octet scalar field,
and $S^A_{R, I}$ are neutral CP-even and CP-odd ones respectively.
In order to avoid tree level FCNC the Yukawa
couplings of these scalars with the SM fermions are usually parameterized as
\cite{cos-Model}
\begin{eqnarray}
{\cal L} = -\eta_{U} Y_{ij}^U \bar{u}_R^i T^A S^A Q_L^j
- \eta_D Y_{ij}^D \bar{d}_R^i T^A (S^A)^\dagger Q_L^j + h.c.,
\end{eqnarray}
where $Y^{U,D}_{ij}$ are the SM Yukawa matrices with $i,j$ denoting flavor
indices, and $\eta_{U,D}$ are flavor universal constants.

The most general renormalizable scalar potential is given by \cite{cos-Model},
\begin{eqnarray}
V&=& \frac{\lambda}{4} \big (H^{\dagger i} H_i - \frac{v^2}{2}\big
)^2 + 2m_S^2 \text{Tr} (S^{\dagger i}S_i) + \lambda_1 H^{\dagger
i}H_i \text{Tr} (S^{\dagger j}S_j) + \lambda_2 H^{\dagger i}H_j
\text{Tr} (S^{\dagger j}S_i)
\nonumber\\
&& + \big [ \lambda_3 H^{\dagger i}
H^{\dagger j} \text{Tr}(S_iS_j) + \lambda_4 H^{\dagger i}
\text{Tr}(S^{\dagger j}S_j S_i) +  \lambda_5 H^{\dagger i}
\text{Tr}(S^{\dagger j}S_i S_j) + h.c. \big ]
\nonumber\\
&& + \lambda_6 \text{Tr}
(S^{\dagger i}S_i S^{\dagger j} S_j) + \lambda_7
\text{Tr}(S^{\dagger i}S_j S^{\dagger j} S_i) + \lambda_8 \text{Tr}
(S^{\dagger i}S_i)\text{Tr}(S^{\dagger j}S_j)\8 \8 \nonumber\\
&& + \lambda_9 \text{Tr}
(S^{\dagger i} S_j) \text{Tr}(S^{\dagger j}S_i) + \lambda_{10}
\text{Tr} (S_i S_j) \text{Tr}(S^{\dagger i}S^{\dagger j})
+\lambda_{11} \text{Tr}(S_iS_j S^{\dagger j} S^{\dagger i}),
\end{eqnarray}
where $S= S^A T^A$ with the color index $A$ summed,  $i,j$ denote isospin indices
and all $\lambda_i$ ($i=1,..., 11$) except $\lambda_4$ and $\lambda_5$ are real parameters \cite{cos-Model}.
Note that by choosing an appropriate phase of the $S$ multiplet, the convention $\lambda_3 >0$ is allowed.
From this potential, one can easily get the mass spectrum of the scalars
\begin{eqnarray}
M_\pm^2 &=& m_S^2 + \lambda_1 \frac{v^2}{4},  \nonumber \\
M_R^2 &=& m_S^2 + (\lambda_1 + \lambda_2 + 2\lambda_3) \frac{v^2}{4}, \nonumber\\
M_I^2 &=& m_S^2 +(\lambda_1 + \lambda_2 - 2\lambda_3) \frac{v^2}{4},
\end{eqnarray}
and their interactions with the color singlet Higgs boson $h$ ($h$ corresponds to the SM Higgs boson) \cite{hgg-NLO}
\begin{eqnarray}
g_{hS^{A\ast}_i S^B_i} = \frac{v}{2} \lambda_i \delta^{AB},
\end{eqnarray}
where $i=\pm, R, I$, and we define $\lambda_\pm = \lambda_1$,
$\lambda_{R,I}=\frac{1}{2}(\lambda_1+\lambda_2 \pm 2 \lambda_3)$.

About the Manohar-Wise model, two points should be noted. One is that just like the $W$ boson in the SM,
$S_\pm$ can contribute to low energy flavor changing processes such as $b \to s \gamma$, and in order to
escape the corresponding experimental constraints, small $|\eta_U \eta_D|$ is favored \cite{cos-Model}.
The importance of $\eta_U$ and $\eta_D$ is that they determine the decay pattern of the scalars, and
consequently, affect their searches at colliders \cite{cos-4j,cos-decay}.
The other is that, although the Yukawa couplings of $h$ with fermions and weak bosons in the model are same as those
of the SM, the couplings of $h$ with gluons, photons and $Z\gamma$ may be changed greatly by the
$S$-mediated loops. Explicitly speaking, in the Manohar-Wise model these couplings are given by \cite{h-scalar}
\begin{eqnarray}
C_{h\gamma\gamma}/SM \equiv \frac{C_{h\gamma\gamma}}{C^{SM}_{h\gamma\gamma}}
  =1+\Big(\frac{C_{h\gamma\gamma}}{C^{SM}_{h\gamma\gamma}} \Big)_{\pm}
  =1+\frac{\frac{2\lambda_1 v^2}{m_{\pm}^{2}}A_0(\tau_{\pm})} {A_1(\tau_W)+\frac{4}{3}A_{\frac{1}{2}}(\tau_t)}, \label{exp1}
\end{eqnarray}
\begin{eqnarray}
C_{hgg}/SM \equiv \frac{C_{hgg}}{C^{SM}_{hgg}}
  =1+\sum_{i=\pm, R, I}\Big(\frac{C_{hgg}}{C^{SM}_{hgg}}\Big)_{i}
  =1+\sum_{i=\pm, R, I}\frac{\frac{3\lambda_i v^2}{4m_{i}^{ 2}}A_0(\tau_{i})}{\frac{1}{2}A_{\frac{1}{2}}(\tau_t)}, \label{exp2}
\end{eqnarray}
\begin{eqnarray}
C_{hZ\gamma}/SM \equiv \frac{C_{hZ\gamma}}{C^{SM}_{hZ\gamma}}
  =1+\Big(\frac{C_{hZ\gamma}}{C^{SM}_{hZ\gamma}}\Big)_{\pm}
  =1-\frac{\frac{2\lambda_1 v^2}{m_{\pm}^{2}} \frac{1-2\sin^2 \theta_W}{\cos\theta_W} C_0(\tau_{\pm}^{-1},\eta_{\pm}^{-1})}
  {C_1(\tau_W^{-1},\eta_W^{-1}) + \frac{2(1-\frac{8}{3}\cos^2\theta_W)}{\cos \theta_W}C_{\frac{1}{2}}(\tau_t^{-1},\eta_t^{-1})},
\end{eqnarray}
where $(C_{hXY}/C^{SM}_{hXY})_i$ with $X,Y=g,\gamma,Z$ denotes $S_i$ ($i=\pm, R, I$) contribution to
the normalized $hXY$ interaction, and $A_0$, $A_{1/2}$, $A_1$, $C_0$, $C_{1/2}$ and
$C_1$ are loop functions defined in \cite{h-rev} with
$\tau_i=m_h^2/4M_i^{2}$ and $\eta_i=m_Z^2/4M_i^{2}$. As a result, the decay width
of $h \to XY$ is now given by \cite{h-cos1}
\begin{eqnarray}
  \Gamma_{h \to \gamma\gamma} =
  \frac{G_{\mu}\alpha^2 m_h^{3}}{128\sqrt{2}\pi^3}
  \Big |A_1(\tau_W)+\frac{4}{3}A_{\frac{1}{2}}(\tau_t)
  +8\times \frac{\lambda_{\pm} v^2}{4m_{\pm}^{2}}A_0(\tau_{\pm}) \Big|^2,
\end{eqnarray}
\begin{eqnarray}
  \Gamma_{h \to gg} =
  \frac{G_{\mu}\alpha^2_{s} m_h^{3}}{16\sqrt{2}\pi^3}
  \Big |\frac{1}{2}A_{\frac{1}{2}}(\tau_t)
  +3\times \sum_{i=\pm, R, I}\frac{\lambda_i v^2}{4m_{i}^{ 2}}A_0(\tau_{i}) \Big|^2,
\end{eqnarray}
\begin{eqnarray}
  \Gamma_{h \to Z\gamma} =
  \frac{G_{\mu}^{2}M_W^{2}\alpha m_h^{3}}{64\pi^4}
  \Big(1-\frac{M_Z^{2}}{M_h^{2}}\Big)^3 \Big |C_1(\tau_W^{-1},\eta_W^{-1})
  +\frac{2(1-\frac{8}{3}\sin^2\theta_W)}{\cos \theta_W} C_{\frac{1}{2}}(\tau_t^{-1},\eta_t^{-1}) \nonumber\\
   -\frac{2 \lambda_{\pm} v^2}{ m_{\pm}^{2}} \frac{1-2\sin^2\theta_W}{\cos\theta_W} C_0(\tau_{\pm}^{-1},\eta_{\pm}^{-1}) \Big |^2.
\end{eqnarray}
Note that our expression for $(C_{hZ\gamma})_\pm$ differs from the formula
in \cite{h-rev} by a minus sign. Such a typo of \cite{h-rev}
was recently pointed out in \cite{Geng}.

Finally, we remind that in the limit $M_\pm \gg v$ and moderate mass splitting of $S_{\pm}$ with $S_{R,I}$,
the forms of the equations (\ref{exp1}) and (\ref{exp2}) can be greatly simplified
\begin{eqnarray}
(C_{hgg}/SM)_{\pm}  &= & -3.6 (C_{h\gamma\gamma}/SM)_{\pm} \simeq 1.149 \times \frac{\lambda_1 v^2}{3 M_{\pm}^{2}}, \label{exp3} \\
(C_{hgg}/SM)_R   &\simeq&  0.575 \times (\lambda_1+\lambda_2+2\lambda_3) \left( \frac{v^2}{3 M_{\pm}^{2}} -
          \frac{(\lambda_2 + 2 \lambda_3)v^4}{12 M_\pm^4} \right) \nonumber \\
                &\simeq & \{ \begin{array}{c} 0.575 \times (\lambda_1 + 4 \lambda_3) ( \frac{v^2}{3 M_\pm^2}
                 -\frac{\lambda_3 v^4}{3 M_\pm^4} )     \quad if\ M_\pm \simeq M_I, \\
                     \frac{1}{2} (C_{hgg}/SM)_\pm \quad \quad \quad \quad \quad \quad \quad \quad if\ M_\pm \simeq M_R,\end{array} \nonumber \\
(C_{hgg}/SM)_I   &\simeq & 0.575 \times (\lambda_1+\lambda_2-2\lambda_3)\left( \frac{v^2}{3 M_{\pm}^{2}} -
          \frac{(\lambda_2 - 2 \lambda_3)v^4}{12 M_\pm^4} \right)   \nonumber \\
                 &\simeq & \{ \begin{array}{c} \frac{1}{2} (C_{hgg}/SM)_\pm \quad \quad \quad \quad \quad \quad \quad  \quad \ if\ M_\pm \simeq M_I, \\
                             0.575 \times (\lambda_1 - 4 \lambda_3) (\frac{v^2}{3 M_\pm^2} + \frac{\lambda_3 v^4}{3 M_\pm^4} )  \quad if\ M_\pm \simeq M_R,
                             \end{array} \nonumber \\
(C_{hgg}/SM ) & = & 1 + 1.149 \times (\frac{\lambda_1 v^2}{3m_{\pm}^{2}}+  \frac{\lambda_R v^2}{3m_{R}^{2}}
        + \frac{\lambda_I v^2}{3m_{I}^{2}}) + \cdots \nonumber \\
      &\simeq &  \{ \begin{array}{c}  1 + 2.3 \times ( \lambda_1+  \lambda_3 ) \frac{v^2}{3m_{\pm}^{2}}  \quad \  if\ M_\pm \simeq M_I, \nonumber \\
       1 + 2.3 \times ( \lambda_1 -  \lambda_3 ) \frac{v^2}{3m_{\pm}^{2}}  \quad \  if\ M_\pm \simeq M_R. \end{array}
\end{eqnarray}
These approximations are very helpful for our later understanding.

\section{Constraints on the Manohar-Wise model}
\subsection{Unitarity Constraint}
In theories with electroweak symmetry breaking,
the unitarity constraint plays an important role in limiting their scalar sector.
This constraint arises from the optical theorem
and it requires the $l$ partial waves in the scattering processes involving
scalars and/or vector bosons satisfy $|a_l| < 1$ \cite{sm-unitarity}.
In actual calculation of pure scalar scattering process
$S_1 S_2 \to S_3 S_4$ in high energy limit, the $J=0$ s-wave amplitude $a_0$  is
approximated by \cite{2hdm-unitarity}
\begin{eqnarray}
a_0 \simeq \frac{1}{16 \pi} Q
\end{eqnarray}
with $Q$ denoting the coupling strength for the four-point vertex
$S_1 S_2 S_3^\ast S_4^\ast$,
and the other partial wave amplitudes are relatively small.
So the unitarity constraint becomes $|Q| < 16 \pi$.
While for the scattering process involving vector bosons, in high energy limit
the dominant contribution come from the longitudinal polarized vector bosons.
And the equivalence theorem states that its amplitude can be approximated by the scalar amplitude
in which the gauge bosons are replaced by their corresponding Goldstone bosons \cite{eqth,eqth-He}.
So the formula for the scalar scattering remains valid in implementing the unitarity constraint.

About the unitarity constraint, another problem one has to face is that the constraint $|a_0| < 1$
is valid for any scattering process $S_i S_j \to S_k S_l$ where $S_i, S_j, S_k$ and $S_l$
represent arbitrary normalized combinations of the scalar fields in the theory, and one must manage
to find the largest value of $|a_0|$ to implement the constraint.
In general, this can be achieved by choosing a set of basis,
such as $\{S_1 S_1, S_1 S_2, \cdots \}$ with $S_i$ denoting the fields in the original Lagrangian,
arraying the s-wave amplitudes for the scatterings
$S_i S_j \to S_k S_l$ with $i,j,k,l=1,2, \cdots$ in matrix form,
and then diagonalizing this matrix to get its eigenvalues \cite{sm-unitarity, 2hdm-unitarity}.
But as far as the Manohar-Wise model is concerned,
such a task is not easy because the model predicts
9 CP-even scalars (i.e. $h$ and $S_R^A$), 9 CP-odd scalars and 18 charged scalars,
and one has to deal with a matrix of $36^2 \times 36^2$ dimension.
Here we point out that since the model preserves electric charge number,
and also keeps CP and SU(3) invariance, one can categorize the basis into
subsets with each of them having definite CP and charge quantum number,
and meanwhile transforming under a certain SU(3) representation.
Considering the transition submatrices based on the subsets
do not couple with each other due to the conservations,
the whole matrix is diagonal in submatrix,
which can greatly simplify the process to find the eigenvalues.
To be more specific, noting the decomposition rule of the tensor product in SU(3) group
\begin{eqnarray}
8 \bigotimes 8 = 1 \bigoplus 8 \bigoplus \bar{8} \bigoplus 10 \bigoplus \overline{10} \bigoplus 27,
\end{eqnarray}
we divide the bi-scalar system (which corresponds to the initial or final state in the scattering)
into 1, 8, 8, 10, 10 and 27 dimension representations respectively.
In the appendix, we present the Clebsch-Gordan coefficients in the decomposition
and the corresponding transition magnitudes for the scattering processes
with the initial and final states lying in a certain SU(3) representation.

In this work, since only $\lambda_1$, $\lambda_2$ and $\lambda_3$ are relevant to our discussion,
we study the unitarity constraint on them by setting other $\lambda_i$ ($i=4, \cdots, 11$) to zero.
For the best-fit value $m_h=125.5$ GeV \cite{1303-c-com,1303-a-com}, we find $|\lambda_1|, |\lambda_2|,
|2 \lambda_1 + \lambda_2| \lesssim 35$ and $\lambda_3 \lesssim 18$. We note that our method can reproduce
the formula in \cite{cos-unitarity-He}, which was obtained ten days later than our work. The difference is in \cite{cos-unitarity-He},
the authors required $Re(a_0)<1/2$, while we required $|a_0|<1$ as in \cite{sm-unitarity}.

\subsection{Collider Constraints}
In Manohar-Wise model, the color-octet scalars are mainly produced in pairs at
hadron colliders \cite{cos-4j}, and experimental efforts to look for them are focused
on dijet-pair events and four-top events. The former
search channel is effective for $\eta_U, \eta_D \simeq 0$. In this case, the scalars are
fermiphobic and at least the lighter neutral scalar will predominantly decay into gluon
pairs through scalar loops \cite{cos-4j}. Then the latest search for dijet-pair events
at 7-TeV LHC, which is performed by the ATLAS collaboration based on $4.6fb^{-1}$ data,
pushes the scalar mass up to above $287{\rm GeV}$ at $95\%$
confidence level \cite{ex-cos-4j}.
Note that such a bound is significantly lower than that
of a color-octet vector boson, which has now been pushed up to about $740~{\rm GeV}$
by the CMS collaboration \cite{ex-cov-4j}. The reason is that the cross section for
scalar pair production process is usually much smaller than that of vector
boson with same mass.  The latter search channel is
pertinent if one of the neutral
scalar dominantly decays into $t\bar{t}$, which can be achieved in the Manohar-Wise
model through a sizable $\eta_U$ \cite{cos-decay}. According to the ATLAS analysis with
about $4.7 fb^{-1}$ data collected at 7-TeV LHC, the measurement
of the same-sign dilepton event rate has put an upper bound on four top quark
production cross section,
which is $61 fb$ at $95\%$ confidence level \cite{ex-4t}. This bound corresponds to
the requirement that the neutral scalar mass must be heavier than about $500 {\rm GeV}$
($630 {\rm GeV}$) if the neutral scalar decays into $t \bar{t}$  at a branching ratio of
$50\%$ ($100\%$). Since all these mass bounds depend on some assumptions, we use a
conservative mass limit of $300{\rm GeV}$ in our discussion.

Maybe the more robust constraint on the exotic scalars comes from electroweak precision data (EWPD).
The dominant way that these scalars influence the electroweak observables,
such as $S$, $T$ and $U$ variables, is though their contributions to the self-energy of the
gauge bosons $\gamma$, $Z$ and $W$ \cite{STU, EWPD}. In this work, we calculate these observables
by the formula presented in \cite{EWPD-fit}, and use the following experimental information to
calculate the corresponding $\chi^2$ \cite{EWPD-fit}:
\begin{eqnarray}
S&=&0.03\pm 0.10, \2
T=0.05\pm 0.12, \2
U=0.03\pm 0.10,
\nonumber \\
M_{STU}&=&
\left(
  \begin{array}{ccc}
     1    &  0.89 & -0.54 \\
     0.89 &  1    & -0.83 \\
    -0.54 & -0.83 &  1 \\
  \end{array}
\right),
\end{eqnarray}
where $M_{STU}$ denotes the correlation coefficient matrix for the three variables. Then we
require $\chi^2 < 8.03$, which corresponds to $95\%$ confidence region defined by the cumulative
distribution function for the three parameter fit, to limit the mass spectrum of the scalars.
We find that the EWPD favor either of the following correlations:
\begin{itemize}
\item $\lambda_2 \simeq 2 \lambda_3$ or equivalently $M_\pm \simeq M_I$.
\item $\lambda_2 \simeq -2 \lambda_3$ or equivalently $M_\pm \simeq M_R$.
\end{itemize}
We note that the former case has been discussed in \cite{EWPD}.

\section{Status of the Manohar-Wise model confronted with the latest Higgs data}

In this section, we perform fits of the model to the latest Higgs data presented at the Rencontres de Moriond 2013 with the method
first proposed in \cite{hfit-egmt, hfit-gkrs} and recently recapitulated in \cite{1212-Gunion}.
These data include the measured signal strengthes for $\gamma \gamma$, $ZZ^\ast$, $W W^\ast$, $b\bar{b}$ and $\tau \bar{\tau}$ channels, and their explicit values are summarized in Fig.2 of \cite{1303-a-com}
(also Fig.6 of this paper) for the ATLAS results, and in Fig.4 of \cite{1303-c-com} for the CMS results.
In our fit, we calculate various observables in Higgs production processes at the LHC with the formula given
in Sect. II and \cite{h-incl-handbook} for fixed $m_h = 125.5$ GeV and $m_t = 173$ GeV, and have properly
considered the correlations of the data like \cite{1212-Gunion}. Noting the fact that, due to
the unknown systematics between the two experiments, the new CMS results in $\gamma \gamma$ channel
($0.78 \pm 0.27$ for mass fit multi-variable analysis and $1.11\pm 0.31$ for cut-based analysis \cite{1303-c-2ph})
are much smaller than their previous results ($1.56^{+0.46}_{-0.42}$ \cite{1211-c-com}) and also than the ATLAS measurement ($1.6\pm 0.3$ \cite{1303-a-com}),
we do not combine the two experimental data together. Instead, we perform two independent fits by
using the ATLAS data and the CMS data respectively.
We conclude that $\chi^2/d.o.f.$
in the SM are $10.55/9$ for the ATLAS data and $4.69/9$ for the CMS data,
and $\chi^2_{min}/d.o.f.$ in the Manohar-Wise model are $5.63/5$ and $2.47/5$ respectively.
Here the total number of d.o.f. is counted in a naive way as $\nu=n_{obs}- n_{para}$, where $n_{obs}$ and $n_{para}$ denote the numbers of the experimental observables and the model free parameters respectively, and for both experiments,
we use  9 sets of data. Note that in the Manohar-Wise model, $\chi^2$ in the SM with the CMS data is much smaller than
that with the ATLAS data, and so is the $\chi^2_{min}$ in the Manohar-Wise model. This is mainly because
for both the collaborations, the same Higgs signal is usually measured from more than one production channels,
and the CMS results are more consistency in the signal rates.
Also note that similar fits with the latest Higgs data have been done
in other new physics frameworks \cite{2hdm-Ferreira, LH-1301, 1302-Cheung}.

\begin{figure}[t]
\centering
\includegraphics[width=16cm]{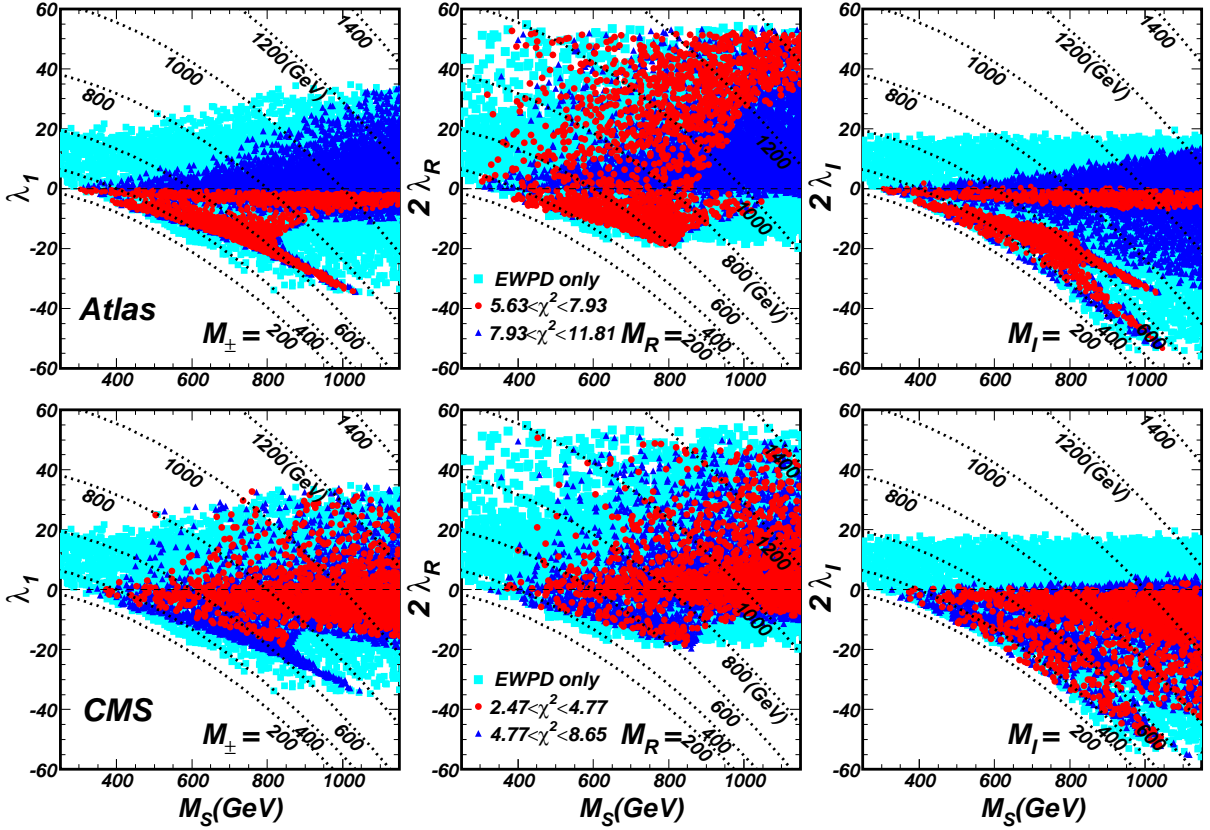}
\vspace*{-0.5cm}
\caption{The scatter plots of the samples surviving different constraints,
projected on the planes
of $\lambda_1$, $2 \lambda_R$($=\lambda_1 + \lambda_2 + 2 \lambda_3$)
and $2 \lambda_I$($= \lambda_1 + \lambda_2 - 2 \lambda_3$) versus $M_S$ respectively.
Here all the samples satisfy the unitarity and collider constraints, while
the red bullets and blue triangles
in the upper (lower) panels represent the samples that can further explain
the ATLAS (CMS) data at $1\sigma$ and $2\sigma$, respectively. }
\label{fig1}
\end{figure}

Our fit procedure is as follows. We first perform an extensive random scan over the following parameter region:
\begin{eqnarray}
300  ~ GeV \leq M_I, M_R, M_\pm \leq 1500 ~ GeV,~
-40 \leq \lambda_1, \lambda_2 \leq 40,~
0 \leq \lambda_3 \leq 20
\end{eqnarray}
We keep the samples satisfying the unitarity constraint and the collider constraints.
Then we calculate the $\chi^2$ associated with each of the surviving samples
and concentrate on two types of them, i.e., those with $\chi^2_{min} \leq \chi^2 \leq \chi^2_{min} + 2.3$
(corresponding to $5.63 \leq \chi^2 \leq 7.93$ for the fit to the ATLAS data and $2.47 \leq \chi^2 \leq 4.77$
for the fit to the CMS data) and those with $\chi^2_{min} + 2.3  < \chi^2 \leq \chi^2_{min} + 6.18$
(corresponding to $7.93 < \chi^2 \leq 11.81$ and $4.77 < \chi^2 \leq 8.65$ respectively).
These two sets of samples correspond to the $68\%$ and $95\%$ confidence level regions
in any two dimensional parameter plane of the model to explain the Higgs data \cite{1212-Gunion},
and hereafter we call them $1\sigma$ and
$2\sigma$ samples respectively.
Note that for most of the $1\sigma$ samples in the fit to the CMS data
and the $2\sigma$ samples in the fit to the ATLAS data, they predict $\chi^2$
smaller than the SM values. This reflects that the
Manohar-Wise model is well suited to explain the current LHC data.

\begin{figure}[t]
\centering
\includegraphics[width=14cm]{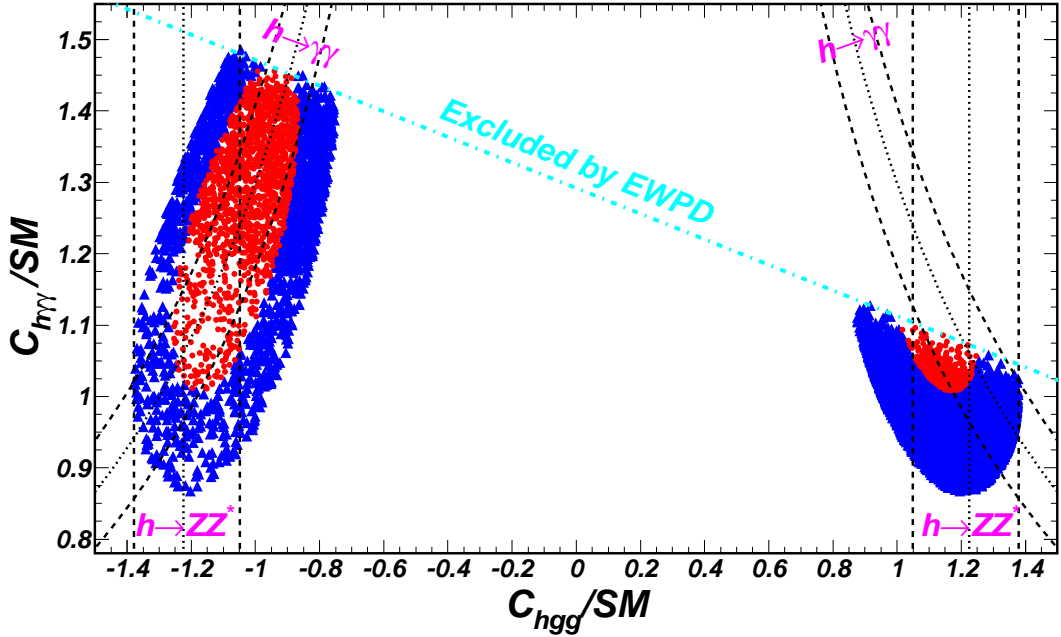}
\vspace*{-0.5cm}
\caption{Same as Fig.\ref{fig1}, but only showing
the samples that can explain the ATLAS data
at $1\sigma$ and $2\sigma$ levels, projected on the plane
of $C_{h\gamma\gamma}/SM$ ($h\gamma \gamma$ coupling
normalized to its SM value) versus $C_{hgg}/SM$
($hgg$ coupling normalized to its SM value).
The central values and the $1\sigma$ regions
for the $\gamma \gamma$ and $ZZ^\ast$ signal rates
from the ATLAS collaboration \cite{1303-a-com}
are also plotted.}
\label{fig2}
\end{figure}

In Fig.\ref{fig1}, we project all the samples that survive the unitarity constraint and the collider constraints on the planes of
$\lambda_1$, $2 \lambda_R $ and $2 \lambda_I $ versus $M_S$ respectively. In order to discriminate the $1\sigma$ and $2 \sigma$
samples from the others, we mark them out with red bullets and blue triangles respectively.  This figure indicates
that the LHC Higgs data are very effective in further limiting the parameter space
which survives the unitarity and collider constraints, especially for the small $M_S$ region.
For example,  at $68\%$ confidence level both the experiments disfavor a positive $\lambda_I$
and the ATLAS data also rule out the possibility of a positive $\lambda_1$.  Another example
is we once counted the number of the $1\sigma$ samples in our random scan,
and we found that in the fit to the ATLAS data, it is only about $0.5\%$ of that for the total samples that satisfy
the unitarity and collider constraints, and in the fit to the CMS data, it is about $7\%$.
This figure also indicates that, due to the great difference of the
$\gamma \gamma$ rate for the two experiments, the parameter space favored by the ATLAS experiment
is quite different from that favored by the CMS experiment. This fact makes it urgent for the two
collaborations to further improve their measurements.
About Fig.\ref{fig1}, we checked that the lower borders of the sky-blue regions for
$M_S \lesssim 950 {\rm GeV}$ are decided by the scalar mass bound, and the other borders are mainly determined
by the unitarity constraint. We also checked that the EWPD prefers the correlation
$\lambda_1 \simeq 2 \lambda_R$ or $\lambda_1 \simeq 2 \lambda_I$.

Considering that the ATLAS data changed little since last June and also that in new physics models it
is difficult to predict a significantly enhanced $\gamma \gamma$ rate relative to its SM prediction,
in the following  we try to illustrate how
the Manohar-Wise model can be used to explain the ATLAS data. We first project in Fig.\ref{fig2} the $1\sigma$ and
$2\sigma$ samples in the upper panels of Fig.\ref{fig1} on the plane of $C_{h\gamma\gamma}/SM$ versus $C_{hgg}/SM$
with $C_{h\gamma\gamma}/SM$ and $C_{hgg}/SM$ denoting the $h\gamma\gamma$ and $hgg$ couplings normalized to their
SM values respectively. In this figure, we also plot the central values (the dot lines) and $1\sigma$ regions
(bounded by the dashed lines) of the $\gamma \gamma$ and $ZZ^\ast$
signals rates measured by the ATLAS collaboration \cite{1303-a-com}.
As expected from the formula of $\chi^2$ \cite{hfit-egmt,1212-Gunion},
for each type of the samples they should form an ellipse \cite{hfit-HSZ}.
We checked that the missing parts of the ellipses are due to the constraints from the EWPD.
Fig.\ref{fig2} exhibits three distinct features. The first feature is that the sign of the $hgg$ coupling
tends to be opposite to that of the SM prediction, especially for the $1\sigma$ samples. The second feature
is the $h\gamma \gamma$ coupling may be enhanced by more than $50\%$, and even if we further require
$\lambda_1, \lambda_R, \lambda_I < 8\pi$ as suggested by the perturbation theory \cite{perturb},
it can still be enhanced by more than $30\%$.  And the last feature is for most samples in Fig.\ref{fig2},
the magnitude of $C_{hgg}/SM$ may exceed unity, which implies an enhanced $ZZ^\ast$ signal at the LHC relative to
its SM expectation. This features is unlikely to be realized in the popular supersymmetric models \cite{susy-Cao, Bhattacharyya}.

\begin{figure}[t]
\centering
\includegraphics[width=9.5cm]{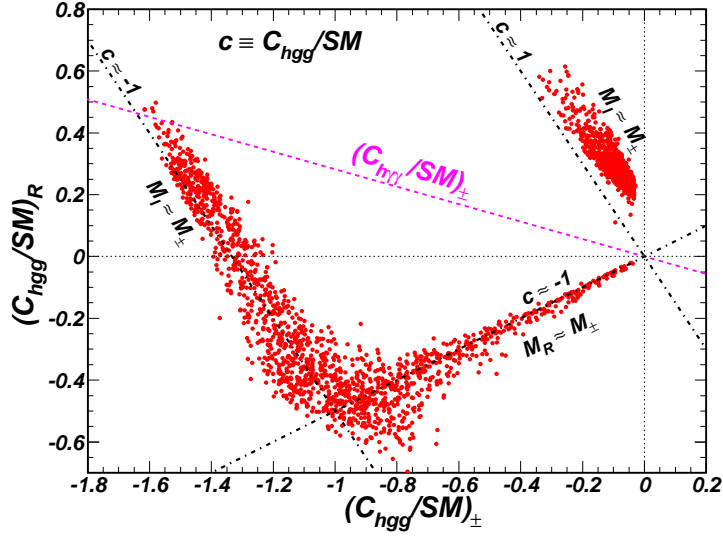}
\vspace*{-0.5cm}
\caption{Same as Fig.\ref{fig2}, but only showing the $1\sigma$ samples, projected
on the plane of $(C_{hgg}/SM)_R$ (contribution of $S_R$)
versus $(C_{hgg}/SM)_\pm$ (contribution of $S_{\pm}$).
The magenta dashed line shows the correlation between $(C_{h\gamma\gamma}/SM)_{\pm}$
and $(C_{hgg}/SM)_{\pm}$ (for this case the vertical axis represents the
$(C_{h\gamma\gamma}/SM)_{\pm}$ value).}
\label{fig3}
\end{figure}

\begin{figure}[htb]
\centering
\includegraphics[width=15cm]{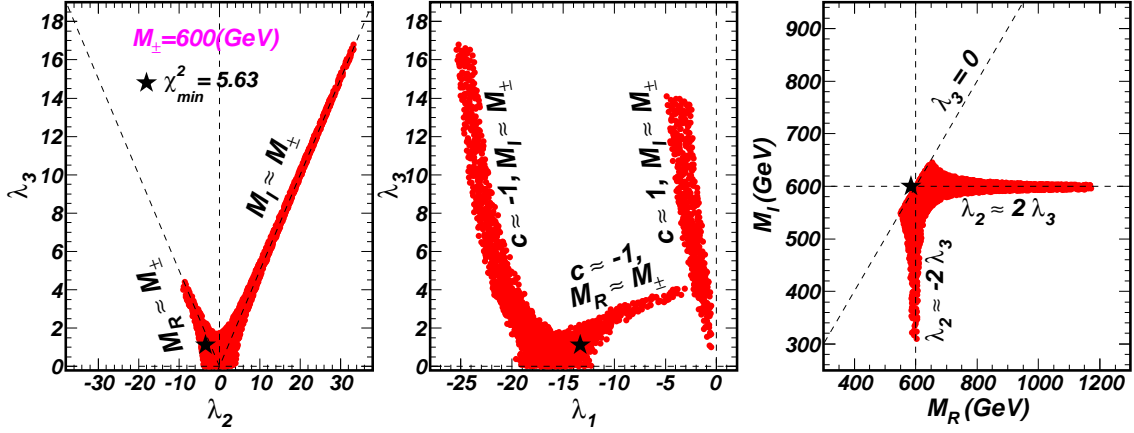}
\vspace*{-0.5cm}
\caption{Same as Fig.\ref{fig3}, but projected on different planes
for $M_{\pm}=600 {\rm ~GeV}$.}
\label{fig4}
\end{figure}

In order to explain these features, we hereafter only consider the $1\sigma$ samples of Fig.\ref{fig2},
and show in Fig.\ref{fig3} the correlation between the $S_\pm$ and $S_R$ contributions to
the $hgg$ coupling. We also fix  $M_\pm = 600 {\rm GeV}$ and show in Fig.\ref{fig4} the
correlations of different input parameters. From these two figures and also the expressions in Eqs.(\ref{exp3}),
one can infer following facts:
\begin{itemize}
\item As shown in the left and right panels of Fig.\ref{fig4}, in the Manohar-Wise model the EWPD prefer
the degeneracy of $M_{\pm}$ with either $M_I$ (corresponding to
  $\lambda_1 = 2 \lambda_I$ or $\lambda_2 = 2 \lambda_3$) or $M_R$ (corresponding to
  $\lambda_1 = 2 \lambda_R$ or $\lambda_2 = -2 \lambda_3$). In the case $M_{\pm} \simeq M_I$,
 \begin{eqnarray}
c \equiv C_{hgg}/SM &=&  1 +  (C_{hgg}/SM)_{\pm} + (C_{hgg}/SM)_{R} + (C_{hgg}/SM)_{I}\nonumber \\
        &\simeq &  1 + 3/2 (C_{hgg}/SM)_{\pm} + (C_{hgg}/SM)_{R} \nonumber \\
        &\simeq &  1 + 2.3 \times ( \lambda_1+  \lambda_3 ) \frac{v^2}{3m_{\pm}^{2}}. \nonumber
\end{eqnarray}
While in the case $M_{\pm} \simeq M_R$,
 \begin{eqnarray}
c \equiv C_{hgg}/SM &=&  1 +  (C_{hgg}/SM)_{\pm} + (C_{hgg}/SM)_{R} + (C_{hgg}/SM)_{I}\nonumber \\
        &\simeq &  1 + 3/2 (C_{hgg}/SM)_{\pm} + (C_{hgg}/SM)_{I} \nonumber \\
        &\simeq &  1 + 2.3 \times ( \lambda_1 -  \lambda_3 ) \frac{v^2}{3m_{\pm}^{2}}. \nonumber
\end{eqnarray}
\item Since the $1\sigma$ samples are characterized by $\lambda_1<0$ and $\lambda_3 > 0 $, we have from Eqs.(\ref{exp3}) that
$(C_{hgg}/SM)_{\pm} < 0$, $(C_{hgg}/SM)_{I} < 0$ and $(C_{h\gamma\gamma}/SM)_{\pm} \simeq - 0.3 (C_{hgg}/SM)_{\pm} > 0$.
As for $(C_{hgg}/SM)_R$ and $C_{hgg}/SM$, they are positive only for the degeneracy
$M_\pm \simeq M_I$ and $\lambda_3 > |\lambda_1|$.
\item For the degeneracy $M_\pm \simeq M_I$, in order to explain the ATLAS data at $1\sigma$ level, $c$ should be around either
$-1$ or $1$ (see Fig.\ref{fig3}).  The former situation occurs for a negatively large $\lambda_1$ (and so is $(C_{hgg}/SM)_\pm$). In this case, the branching ratio for $h \to \gamma \gamma$ is greatly enhanced (see Fig.\ref{fig3}),
and meanwhile the $hgg$ coupling can be well tuned by $\lambda_3$ (see the middle panel of Fig.\ref{fig4}).
As a result, a rather low $\chi^2$ can be obtained. While for the situation of $c \simeq 1$, it occurs only for a small $|\lambda_1|$. Consequently the branching ratio for $h \to \gamma \gamma$ changes little,
and $\chi^2$ is usually large.
\item For the degeneracy $M_\pm \simeq M_R$, $c$ should be around -1. Since in this case, all scalar contributions
to the $hgg$ coupling are negative, the parameter $\lambda_3$ is not necessarily very large (see the middle panel of Fig.\ref{fig4}).
\item At the turning point where the degeneracy $M_\pm \simeq M_R$ converts to $M_\pm \simeq M_I$,
$\lambda_3 = \lambda_2 =0$. So as $\lambda_1$ becomes negatively larger from zero point, $\lambda_3$ first
decreases before reaching the turning point, then increases monotonously in departing the point (see middle panel of
Fig.\ref{fig4}). We numerically checked that this is true for $M_{\pm} \lesssim 700 {\rm GeV}$. While for
$M_{\pm} \gtrsim 700 {\rm GeV}$, the unitarity requires $(\lambda_1 - \lambda_3) \gtrsim -17$ for the degeneracy
 $M_\pm \simeq M_R$, which implies that $C_{h\gamma\gamma}/SM \lesssim 1- 0.013\times\lambda_1 \lesssim 1.22$
and $C_{hgg}/SM \gtrsim 1 + 0.1\times(\lambda_1-\lambda_3) \gtrsim -0.7$. In such situation,
the tuning point can not be used to explain the ATLAS data at $1 \sigma$ level, and that is why $\lambda_1$ is located
in two separated regions (see upper left panel of Fig.\ref{fig1}).
\end{itemize}

\begin{table}
\caption{Detailed information for some benchmark points in the Manohar-Wise model.}
\renewcommand{\arraystretch}{0.92}
  \setlength{\tabcolsep}{15pt}
  \centering
\begin{tabular}{|c|c|c|c|}
  \hline \hline
  Benchmark Point & P1 & P2 & P3 \\
  \hline
  $\chi^2$   &   5.63 &   5.63 &   6.28 \\
  $M_{\pm}(GeV)$  &  400.00 &  600.00 &  800.00 \\
  $M_R(GeV)$ &  404.76 &  584.21 &  804.41 \\
  $M_I(GeV)$ &  337.59 & 523.32 & 512.31  \\
  \hline
  $M_S(GeV)$ &  498.38 &  749.93 &  901.78 \\
  $\lambda_1$ &  -5.83 &  -13.35 &   -11.43    \\
  $\lambda_2$ &   -1.39 &   -3.46 &   -12.22     \\
  $\lambda_3$ &    0.82 &   1.11 &   6.34   \\
  \hline
  $C_{h\gamma\gamma}/SM$ &   1.230 &    1.232 &   1.111 \\
  $C_{hgg}/SM$ &   -1.066 &   -1.067 &    -1.119\\
  $(C_{hZ\gamma}/SM)_{\pm}$ &   0.085 &   0.086 &   0.041\\
  $(C_{hgg}/SM)_{\pm}$ &   -0.813 &   -0.822  &   -0.395 \\
  $(C_{hgg}/SM)_R$ &    -0.380 &    -0.474 &    -0.187 \\
  $(C_{hgg}/SM)_I$ &   -0.873 &   -0.771 &   -1.537 \\
  \hline
     LHC, ggF+ttH, $\gamma\gamma$
     & 1.697 & 1.705 & 1.512
     \\
     LHC, VBF+VH, $\gamma\gamma$
     & 1.705 & 1.498 & 1.209
     \\
     LHC, ggF+ttH, $ZZ^*$
     & 1.494 & 1.123 & 1.224
     \\
     LHC, VBF+VH, $ZZ^*$
     & 1.127 & 0.987 & 0.979
     \\
     LHC, ggF+ttH, $WW^*$
     & 0.987 & 1.123 & 1.224
     \\
     LHC, VBF+VH, $WW^*$
     & 1.127 & 0.987 & 0.979
     \\
     LHC,VH tag, $b\bar{b}$
     & 0.987 & 0.987 & 0.979
     \\
     LHC, ggF+ttH, $\tau\tau$
     & 0.987 & 1.123 & 1.224
     \\
     LHC, VBF+VH, $\tau\tau$
     & 1.127 & 0.987 & 0.979
     \\
  \hline
  \hline
\end{tabular}
\label{tab1}\vspace{-0.35cm}
\end{table}

\begin{figure}[t]
\centering
\includegraphics[width=12cm]{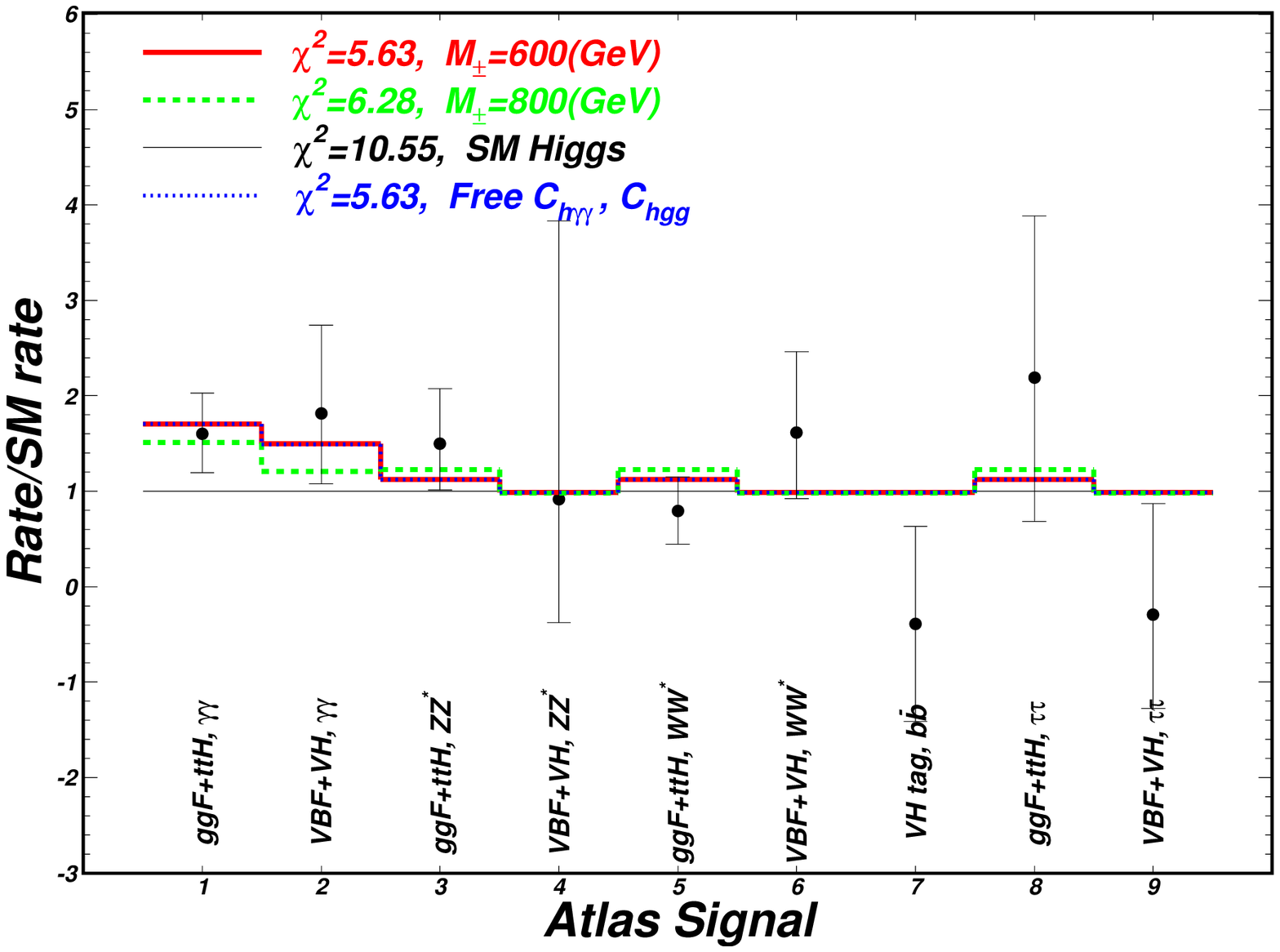}
\vspace*{-0.5cm}
\caption{Predictions of various Higgs signal rates for the second and third benchmark points
presented in Table I, compared with the ATLAS data \cite{1303-a-com}.
The results given by the best point with
free $C_{hgg}$ and $C_{h\gamma\gamma}$ couplings are also shown.}
\label{fig5}
\end{figure}

As a completion of Fig.\ref{fig2}, we also present the details of the best-fit points in Table \ref{tab1},
and compare their predictions on different Higgs observables with the corresponding experimental
data in Fig.\ref{fig5}. This figure indicates that for the best point, most of its theoretical
predictions agree with the experimental data at $1\sigma$ level, and the best
explanations are achieved for the ATLAS results in $\gamma\gamma$ channels.

\begin{figure}[t]
\centering
\includegraphics[width=12cm]{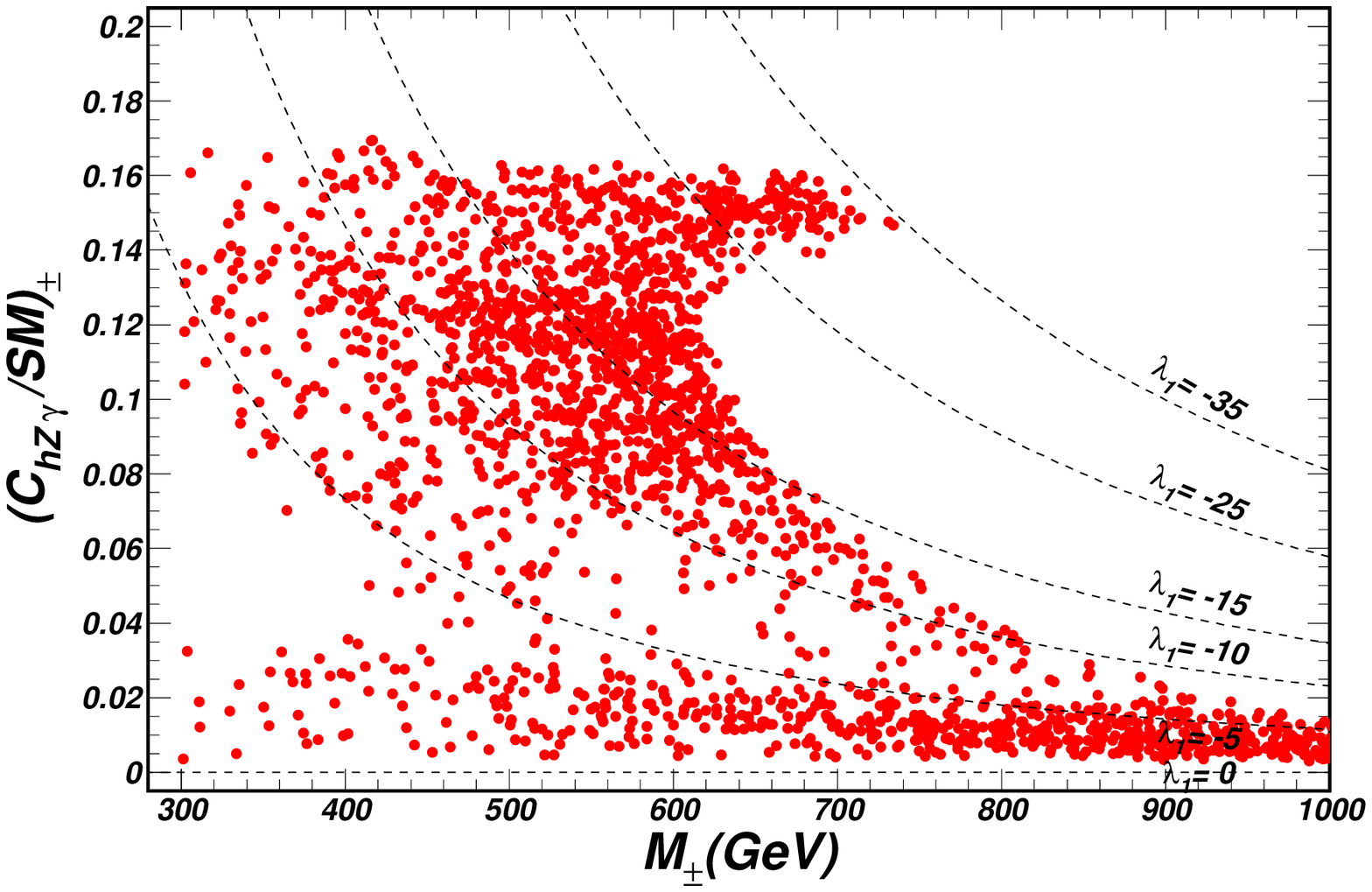}
\vspace*{-0.5cm}
\caption{Same as Fig.\ref{fig3},
but showing the relative correction of the Manohar-Wise model
to $hZ\gamma$ coupling. Note that such a correction only comes from
the $S_\pm$ mediated loops.}
\label{fig6}
\end{figure}
Noting that the decay $h \to Z \gamma$ was recently investigated both experimentally \cite{1303-a-zg, 1303-c-zg}
and theoretically \cite{Geng, hsusy-zg}, we also examine the $hZ\gamma$ coupling in this work.
In the Manohar-Wise model,
this coupling receives new correction only from the $S_\pm$ mediated loops,
so the correction size depends on $M_\pm$ and $\lambda_1$.
In Fig.\ref{fig5} we show such dependences.
This figure indicates that in contrast with possible large correction of $S_\pm$ to the $h\gamma\gamma$ coupling,
the $S_\pm$ correction to the $hZ\gamma$ coupling can only reach $17\%$.
The reason is that the $ZS^+S^-$ coupling strength is relatively small, i.e.,
$C_{ZS^+S^-} \simeq 0.3 C_{\gamma S^+S^-}$.

Like Fig.\ref{fig2}, \ref{fig3} and \ref{fig4}, one may also investigate the features of the Manohar-Wise model in
explaining the CMS data, but now the situation is quite complicated since for the $1\sigma$ samples in the bottom panels
of Fig.\ref{fig1}, $\lambda_1$ may be either positive or negative. On the other hand, considering that the CMS
data in 2013 is quite different from those in 2012, we incline
to wait for new data of the CMS collaboration before doing this.

\section{Conclusions}
Since the discovery of the Higgs-like boson at the LHC,
the experimental precision to determine its properties has been
improved significantly as more and more data are accumulated.
In light of this progress
in experiment, it is urgent for theorists to investigate
how well the new physics models
can explain the data and what is the key feature of the models
in doing this. In this work, we try to answer
these questions in the Manohar-Wise model.
As the first step of our research, we examine the constraints
on the model, which include the unitarity constraint, the collider searches
for new scalars and the EWPD.
In implementing the unitarity constraint, we note it is a very complicated
work in the Manohar-Wise model and has not been considered before,
so we have to treat it in a special way. With the help of some knowledge
of group theory, we succeed in solving such a problem.
Next we perform a fit of the model by building
an appropriate $\chi^2$ function. Our fit procedure is as follows. We scan the parameter space of the
model and only retain the samples that satisfy various constraints.
Then with the latest Higgs data released at the Rencontres de Moriond 2013,
we calculate the $\chi^2$ value associated with each of the surviving samples,
and determine the $68\%$ and $95\%$ confidence level regions in any two dimensional parameter
plane of the model to explain the Higgs data. In the calculation, we perform fits by employing the data
from the ATLAS collaboration and from the CMS collaboration separately since due to unknown systematics,
the measured $\gamma\gamma$ and  $ZZ^*$ rates of the two collaborations are quite different.
Considering that in new physics models, it is difficult to predict significantly enhanced
$\gamma \gamma$ rate, we also illustrate how the Manohar-Wise model is capable in doing this.

Base on our analysis, we have following conclusions:
\begin{itemize}
\item The Manohar-Wise model is able to explain the ATLAS data and the CMS data quite well,
with the resulting $\chi^2$ significantly smaller than its corresponding SM $\chi^2$.
In particular, in order to explain the $\gamma \gamma$ enhancement reported by the ATLAS collaboration,
the sign of the $hgg$ coupling is usually opposite to that in the SM.
\item Current Higgs data, especially the ATLAS data, are very powerful in further limiting
 the parameter space of the model that satisfies the unitary and collider constraints.
 After considering all the constraints, the degeneracy $M_\pm \simeq M_I$
 or $M_\pm \simeq M_R$ is strongly preferred, and $\lambda_1$ is required to be less than 5 for
 $M_S \lesssim 500 {\rm GeV}$.
\end{itemize}

\section*{Acknowledgement}
We thank Lei Wu, Peiwen Wu and Yang Zhang for helpful discussion.
This work was supported in part by the National Natural
Science Foundation of China (NNSFC) under grant  No. 10775039, 11075045,
11275245, 10821504 and 11135003, by Specialized Research Fund for
the Doctoral Program of Higher Education with grant No. 20104104110001,
and  by the Project of Knowledge
Innovation Program (PKIP) of Chinese Academy of Sciences under grant
No. KJCX2.YW.W10.

\section*{Appendix A}

In our method, the $\alpha$th state ($\alpha=1,2,\cdots, N$) of the bi-color-octet-scalar system
(corresponding to the initial or final state in the scattering)
in N-dimensional representation of SU(3) group is written as
\begin{eqnarray}
B^N[\alpha] = \sum_{A,B=1}^8 S^A (C^N[\alpha])_{AB} S^B,
\end{eqnarray}
where $C^N[\alpha]$ denotes the Clebsch-Gordan coefficient for the decomposition in matrix form. Then in
the following when we talk about the basis $(B_1^N,B_2^N,\cdots)$, it actually represents the collection of states
$(B_1^N[1],B_1^N[2],\cdots,B_1^N[N],B_2^N[1],B_2^N[2],\cdots)$. But on the hand, because $SU(3)$ symmetry is
unbroken in Manohar-Wise model, the transition amplitude for the scattering process
$B_i^N [\alpha] \to B_j^{N^\prime} [\beta]$ must take the form $A_{ij} \delta_{N N^\prime} \delta_{\alpha \beta}$,
so without loss of any information, we can neglect the color index $\alpha$ of the basis and only investigate the dependence of
$A_{ij}$ on model parameters. In the following, in order to present our formula in a neat way we define following abbreviations
\begin{eqnarray}
  \lambda_{1,2}^{+} = \lambda_1+\lambda_2, \nonumber \\
  \lambda_{6,7}^{+} = \lambda_6+\lambda_7, \nonumber \\
  \lambda_{8,9}^{+}=\lambda_8+\lambda_9, \nonumber \\
  \lambda_{6,7,11}^{++}=\lambda_6+\lambda_7+\lambda_{11}, \nonumber
\end{eqnarray}
and only list the expression of non-zero scattering amplitudes.

1. Scatterings between states in 1-dimensional representation of the SU(3) group.

In this case, the Clebsch-Gordan coefficients are given by
\begin{eqnarray}
{ C^1[1]}=\left(
\begin{array}{cccccccc}
\frac{\sqrt{2}}{4}&    0&    0&    0&    0&    0&    0&    0\\
0&    \frac{\sqrt{2}}{4}&    0&    0&    0&    0&    0&    0\\
0&    0&    \frac{\sqrt{2}}{4}&    0&    0&    0&    0&    0\\
0&    0&    0&    \frac{\sqrt{2}}{4}&    0&    0&    0&    0\\
0&    0&    0&    0&    \frac{\sqrt{2}}{4}&    0&    0&    0\\
0&    0&    0&    0&    0&    \frac{\sqrt{2}}{4}&    0&    0\\
0&    0&    0&    0&    0&    0&    \frac{\sqrt{2}}{4}&    0\\
0&    0&    0&    0&    0&    0&    0&    \frac{\sqrt{2}}{4}\\
\end{array}
\right). \nonumber
\end{eqnarray}

\begin{itemize}
\item Denoting $A_{ij}$ as the S-partial wave amplitude for the scattering
$B_i \to B_j$, with $B_i$ and $B_j$ representing any state in the basis
\{ $\omega^-\omega^+$, $\frac{zz}{\sqrt{2}}$, $\frac{hh}{\sqrt{2}}$,
$(S^-S^+)^1$, $\frac{(S_IS_I)^1}{\sqrt{2}}$, $\frac{(S_RS_R)^1}{\sqrt{2}}$ \}, we have \\
 $A_{11} = \lambda$, \quad
 $A_{12} = A_{13} =\frac{\sqrt{2}}{4}\lambda,$ \quad
 $A_{14} =  \sqrt{2}\lambda_{1,2}^{+}$,\quad
 $A_{15} = A_{16} = A_{24} = A_{34} = \lambda_{1}$,\\
 $A_{22} = A_{33} = \frac{3}{4}\lambda$, \quad
 $A_{23}  = \frac{1}{4}\lambda$, \quad
 $A_{25} = A_{36} = \sqrt{2}(\frac{1}{2}\lambda_{1,2}^{+}+\lambda_{3})$,\\
 $A_{26} = A_{35} = \sqrt{2}(\frac{1}{2}\lambda_{1,2}^{+}-\lambda_{3})$,\quad
 $A_{44} = \frac{8}{3}\lambda_{6,7}^{+}+\frac{9}{2}\lambda_{8,9}^{+}
          +\lambda_{10}+\frac{7}{6}\lambda_{11}$,\\
 $A_{45} = A_{46} = \frac{1}{\sqrt{2}}(\frac{4}{3}\lambda_{6,7,11}^{++}
          +\frac{1}{2}\lambda_{9,10}^{+}+4\lambda_{8})$,\quad
 $A_{55} = A_{66} = \frac{1}{2}(\frac{5}{2}\lambda_{6,7,11}^{++}
          +5\lambda_{8,9,10}^{++})$,\\
 $A_{56} = \frac{1}{2}(\frac{17}{6}\lambda_{6,7}^{+}+4\lambda_{8,9}^{+}
          -3\lambda_{10}-\frac{1}{6}\lambda_{11})$.

\item In the basis \{ $hz$, $(S_IS_R)^1$ \}, $A_{ij}$ is given by \\
 $A_{11} = \frac{1}{2}\lambda$,\quad
 $A_{12} = 2\sqrt{2}\lambda_{3}$,\quad
 $A_{22} = -\frac{1}{6}\lambda_{6,7}^{+}+\frac{1}{2}\lambda_{8,9}^{+}
           +4\lambda_{10}+\frac{4}{3}\lambda_{11}$.

 \item
In the basis \{ $h\omega^-$, $z\omega^-$, $(S_RS^-)^1$, $(S_IS^-)^1$ \},
$A_{ij}$ is given by \\
$A_{11} = A_{22} = \frac{1}{2}\lambda$,\quad
$A_{12} = 0$,\quad
$A_{13} = A_{24} =  \sqrt{2}(\frac{1}{2}\lambda_{2}+\lambda_{3})$,\quad
$A_{34} =-\frac{3}{4}i\lambda_{6,7,11}^{-+}+\frac{7}{4}i\lambda_{9,10}^{-}$,\\
$A_{14} = -A_{23} = \sqrt{2}i(\frac{1}{2}\lambda_{2}-\lambda_{3})$,\quad
$A_{33} = A_{44} = \frac{7}{12}\lambda_{6,7,11}^{++}
          +\frac{9}{4}\lambda_{9,10}^{+}+\frac{1}{2}\lambda_{8}$.

\end{itemize}

2. Scatterings between states in 8-dimensional representation of the SU(3) group.

In this case, the Clebsch-Gordan coefficients for 8 representation are proportional to
$d^{\alpha \beta \gamma}$, and those for $\bar{8}$ representation are proportional to
$f^{\alpha \beta \gamma}$. The values of these coefficients are
\begin{eqnarray}
{ C^8[1]}=\left(
\begin{array}{cccccccc}
0&    0&    0&    0&    0&    0&    0&    \frac{\sqrt{10}}{10}\\
0&    0&    0&    0&    0&    0&    0&    \frac{\sqrt{10}}{10}i \\
0&    0&    0&    0&    0&    0&    0&    0\\
0&    0&    0&    0&    0&    \frac{\sqrt{30}}{20}&   -\frac{\sqrt{30}}{20}i &   0\\
0&    0&    0&    0&    0&    \frac{\sqrt{30}}{20}i &   \frac{\sqrt{30}}{20}&   0\\
0&    0&    0&    \frac{\sqrt{30}}{20}&   \frac{\sqrt{30}}{20}i &   0&    0&    0\\
0&    0&    0&    -\frac{\sqrt{30}}{20}i &   \frac{\sqrt{30}}{20}&   0&    0&    0\\
\frac{\sqrt{10}}{10}&   \frac{\sqrt{10}}{10}i &   0&    0&    0&    0&    0&    0\\
\end{array}\right)\nonumber
\end{eqnarray}

\begin{eqnarray}
{ C^8[2]}=\left(
\begin{array}{cccccccc}
0&    0&    0&    0&    0&    -\frac{\sqrt{30}}{20}&  -\frac{\sqrt{30}}{20}i &
0\\
0&    0&    0&    0&    0&    -\frac{\sqrt{30}}{20}i &   \frac{\sqrt{30}}{20}&   0\\
0&    0&    0&    -\frac{\sqrt{30}}{20}&  -\frac{\sqrt{30}}{20}i &   0&    0&    0\\
0&    0&    -\frac{\sqrt{30}}{20}&  0&    0&    0&    0&    \frac{\sqrt{10}}{20}\\
0&    0&    -\frac{\sqrt{30}}{20}i &   0&    0&    0&    0&    \frac{\sqrt{10}}{20}i \\
-\frac{\sqrt{30}}{20}&  -\frac{\sqrt{30}}{20}i &   0&    0&    0&    0&    0&    0\\
-\frac{\sqrt{30}}{20}i &   \frac{\sqrt{30}}{20}&   0&    0&    0&    0&    0&    0\\
0&    0&    0&    \frac{\sqrt{10}}{20}&   \frac{\sqrt{10}}{20}i &   0&    0&    0\\
\end{array}\right)\nonumber
\end{eqnarray}

\begin{eqnarray}
{ C^8[3]}=\left(
\begin{array}{cccccccc}
\frac{\sqrt{15}}{10}&   0&    0&    0&    0&    0&    0&    0\\
0&    \frac{\sqrt{15}}{10}&   0&    0&    0&    0&    0&    0\\
0&    0&    \frac{\sqrt{15}}{10}&   0&    0&    0&    0&    \frac{\sqrt{5}}{10}\\
0&    0&    0&    0&    0&    0&    0&    0\\
0&    0&    0&    0&    0&    0&    0&    0\\
0&    0&    0&    0&    0&    -\frac{\sqrt{15}}{10}&  0&    0\\
0&    0&    0&    0&    0&    0&    -\frac{\sqrt{15}}{10}&  0\\
0&    0&    \frac{\sqrt{5}}{10}&   0&    0&    0&    0&    -\frac{\sqrt{15}}{10}\\
\end{array}\right)\nonumber
\end{eqnarray}

\begin{eqnarray}
{ C^8[4]}=\left(
\begin{array}{cccccccc}
0&    0&    0&    0&    0&    \frac{\sqrt{30}}{20}&   -\frac{\sqrt{30}}{20}i &
0\\
0&    0&    0&    0&    0&    -\frac{\sqrt{30}}{20}i &   -\frac{\sqrt{30}}{20}&  0\\
0&    0&    0&    \frac{\sqrt{30}}{20}&   -\frac{\sqrt{30}}{20}i &   0&    0&    0\\
0&    0&    \frac{\sqrt{30}}{20}&   0&    0&    0&    0&    -\frac{\sqrt{10}}{20}\\
0&    0&    -\frac{\sqrt{30}}{20}i &   0&    0&    0&    0&    \frac{\sqrt{10}}{20}i \\
\frac{\sqrt{30}}{20}&   -\frac{\sqrt{30}}{20}i &   0&    0&    0&    0&    0&    0\\
-\frac{\sqrt{30}}{20}i &   -\frac{\sqrt{30}}{20}&  0&    0&    0&    0&    0&    0\\
0&    0&    0&    -\frac{\sqrt{10}}{20}&  \frac{\sqrt{10}}{20}i &   0&    0&    0\\
\end{array}\right)\nonumber
\end{eqnarray}

\begin{eqnarray}
{ C^8[5]}=\left(
\begin{array}{cccccccc}
0&    0&    0&    0&    0&    0&    0&     \frac{\sqrt{10}}{10}\\
0&    0&    0&    0&    0&    0&    0&    -\frac{\sqrt{10}}{10}i \\
0&    0&    0&    0&    0&    0&    0&    0\\
0&    0&    0&    0&    0&    \frac{\sqrt{30}}{20}&   \frac{\sqrt{30}}{20}i &   0\\
0&    0&    0&    0&    0&    -\frac{\sqrt{30}}{20}i &   \frac{\sqrt{30}}{20}&   0\\
0&    0&    0&    \frac{\sqrt{30}}{20}&   -\frac{\sqrt{30}}{20}i &   0&    0&    0\\
0&    0&    0&    \frac{\sqrt{30}}{20}i &   \frac{\sqrt{30}}{20}&   0&    0&    0\\
\frac{\sqrt{10}}{10}&   -\frac{\sqrt{10}}{10}i &   0&    0&    0&    0&    0&    0\\
\end{array}\right)\nonumber
\end{eqnarray}

\begin{eqnarray}
{ C^8[6]}=\left(
\begin{array}{cccccccc}
0&    0&    0&    \frac{\sqrt{30}}{20}&   -\frac{\sqrt{30}}{20}i &   0&    0&
0\\
0&    0&    0&    \frac{\sqrt{30}}{20}i &   \frac{\sqrt{30}}{20}&   0&    0&    0\\
0&    0&    0&    0&    0&    -\frac{\sqrt{30}}{20}&  \frac{\sqrt{30}}{20}i &   0\\
\frac{\sqrt{30}}{20}&   \frac{\sqrt{30}}{20}i &   0&    0&    0&    0&    0&    0\\
-\frac{\sqrt{30}}{20}i &   \frac{\sqrt{30}}{20}&   0&    0&    0&    0&    0&    0\\
0&    0&    -\frac{\sqrt{30}}{20}&  0&    0&    0&    0&    -\frac{\sqrt{10}}{20}\\
0&    0&    \frac{\sqrt{30}}{20}i &   0&    0&    0&    0&    \frac{\sqrt{10}}{20}i \\
0&    0&    0&    0&    0&    -\frac{\sqrt{10}}{20}&  \frac{\sqrt{10}}{20}i &   0\\
\end{array}\right)\nonumber
\end{eqnarray}

\begin{eqnarray}
{ C^8[7]}=\left(
\begin{array}{cccccccc}
0&    0&    0&    -\frac{\sqrt{30}}{20}&  -\frac{\sqrt{30}}{20}i &   0&    0&
0\\
0&    0&    0&    \frac{\sqrt{30}}{20}i &   -\frac{\sqrt{30}}{20}&  0&    0&    0\\
0&    0&    0&    0&    0&    \frac{\sqrt{30}}{20}&   \frac{\sqrt{30}}{20}i &   0\\
-\frac{\sqrt{30}}{20}&  \frac{\sqrt{30}}{20}i &   0&    0&    0&    0&    0&    0\\
-\frac{\sqrt{30}}{20}i &   -\frac{\sqrt{30}}{20}&  0&    0&    0&    0&    0&    0\\
0&    0&    \frac{\sqrt{30}}{20}&   0&    0&    0&    0&    \frac{\sqrt{10}}{20}\\
0&    0&    \frac{\sqrt{30}}{20}i &   0&    0&    0&    0&    \frac{\sqrt{10}}{20}i \\
0&    0&    0&    0&    0&    \frac{\sqrt{10}}{20}&   \frac{\sqrt{10}}{20}i &   0\\
\end{array}\right)\nonumber
\end{eqnarray}

\begin{eqnarray}
{ C^8[8]}=\left(
\begin{array}{cccccccc}
\frac{\sqrt{5}}{10}&   0&    0&    0&    0&    0&    0&    0\\
0&    \frac{\sqrt{5}}{10}&   0&    0&    0&    0&    0&    0\\
0&    0&    \frac{\sqrt{5}}{10}&   0&    0&    0&    0&    -\frac{\sqrt{15}}{10}\\
0&    0&    0&    -\frac{\sqrt{5}}{5}&  0&    0&    0&    0\\
0&    0&    0&    0&    -\frac{\sqrt{5}}{5}&  0&    0&    0\\
0&    0&    0&    0&    0&    \frac{\sqrt{5}}{10}&   0&    0\\
0&    0&    0&    0&    0&    0&    \frac{\sqrt{5}}{10}&   0\\
0&    0&    -\frac{\sqrt{15}}{10}&  0&    0&    0&    0&    -\frac{\sqrt{5}}{10}\\
\end{array}\right)\nonumber
\end{eqnarray}

\begin{eqnarray}
{ C^{\bar{8}}[1]}=\left(
\begin{array}{cccccccc}
0&    0&    \frac{\sqrt{6}}{6}&   0&    0&    0&    0&    0\\
0&    0&    \frac{\sqrt{6}}{6}i &   0&    0&    0&    0&    0\\
-\frac{\sqrt{6}}{6}&  -\frac{\sqrt{6}}{6}i &   0&    0&    0&    0&    0&    0\\
0&    0&    0&    0&    0&    -\frac{\sqrt{6}}{12}&  \frac{\sqrt{6}}{12}i &   0\\
0&    0&    0&    0&    0&    -\frac{\sqrt{6}}{12}i &   -\frac{\sqrt{6}}{12}&  0\\
0&    0&    0&    \frac{\sqrt{6}}{12}&   \frac{\sqrt{6}}{12}i &   0&    0&    0\\
0&    0&    0&    -\frac{\sqrt{6}}{12}i &   \frac{\sqrt{6}}{12}&   0&    0&    0\\
0&    0&    0&    0&    0&    0&    0&    0\\
\end{array}\right)\nonumber
\end{eqnarray}

\begin{eqnarray}
{ C^{\bar{8}}[2]}=\left(
\begin{array}{cccccccc}
0&    0&    0&    0&    0&    \frac{\sqrt{6}}{12}&   \frac{\sqrt{6}}{12}i & 0\\
0&    0&    0&    0&    0&    \frac{\sqrt{6}}{12}i &   -\frac{\sqrt{6}}{12}&  0\\
0&    0&    0&    \frac{\sqrt{6}}{12}&   \frac{\sqrt{6}}{12}i &   0&    0&    0\\
0&    0&    -\frac{\sqrt{6}}{12}&  0&    0&    0&    0&    -\frac{\sqrt{2}}{4}\\
0&    0&    -\frac{\sqrt{6}}{12}i &   0&    0&    0&    0&    -\frac{\sqrt{2}}{4}i \\
-\frac{\sqrt{6}}{12}&  -\frac{\sqrt{6}}{12}i &   0&    0&    0&    0&    0&    0\\
-\frac{\sqrt{6}}{12}i &   \frac{\sqrt{6}}{12}&   0&    0&    0&    0&    0&    0\\
0&    0&    0&    \frac{\sqrt{2}}{4}&   \frac{\sqrt{2}}{4}i &   0&    0&    0\\
\end{array}\right)\nonumber
\end{eqnarray}

\begin{eqnarray}
{ C^{\bar{8}}[3]}=\left(
\begin{array}{cccccccc}
0&    \frac{\sqrt{3}}{6}i &   0&    0&    0&    0&    0&    0\\
-\frac{\sqrt{3}}{6}i &   0&    0&    0&    0&    0&    0&    0\\
0&    0&    0&    0&    0&    0&    0&    0\\
0&    0&    0&    0&    \frac{\sqrt{3}}{3}i &   0&    0&    0\\
0&    0&    0&    -\frac{\sqrt{3}}{3}i &   0&    0&    0&    0\\
0&    0&    0&    0&    0&    0&    \frac{\sqrt{3}}{6}i &   0\\
0&    0&    0&    0&    0&    -\frac{\sqrt{3}}{6}i &   0&    0\\
0&    0&    0&    0&    0&    0&    0&    0\\
\end{array}\right)\nonumber
\end{eqnarray}

\begin{eqnarray}
{ C^{\bar{8}}[4]}=\left(
\begin{array}{cccccccc}
0&    0&    0&    0&    0&    \frac{\sqrt{6}}{12}&
-\frac{\sqrt{6}}{12}i &0\\
0&    0&    0&    0&    0&    -\frac{\sqrt{6}}{12}i &   -\frac{\sqrt{6}}{12}&  0\\
0&    0&    0&    \frac{\sqrt{6}}{12}&   -\frac{\sqrt{6}}{12}i &   0&    0&    0\\
0&    0&    -\frac{\sqrt{6}}{12}&  0&    0&    0&    0&    -\frac{\sqrt{2}}{4}\\
0&    0&    \frac{\sqrt{6}}{12}i &   0&    0&    0&    0&    \frac{\sqrt{2}}{4}i \\
-\frac{\sqrt{6}}{12}&  \frac{\sqrt{6}}{12}i &   0&    0&    0&    0&    0&    0\\
\frac{\sqrt{6}}{12}i &   \frac{\sqrt{6}}{12}&   0&    0&    0&    0&    0&    0\\
0&    0&    0&    \frac{\sqrt{2}}{4}&   -\frac{\sqrt{2}}{4}i &   0&    0&
0\\
\end{array}\right)\nonumber
\end{eqnarray}

\begin{eqnarray}
{ C^{\bar{8}}[5]}=\left(
\begin{array}{cccccccc}
0&    0&    -\frac{\sqrt{6}}{6}&  0&    0&    0&    0&   0\\
0&    0&    \frac{\sqrt{6}}{6}i &   0&    0&    0&    0&    0\\
\frac{\sqrt{6}}{6}&   -\frac{\sqrt{6}}{6}i &   0&    0&    0&    0&    0&    0\\
0&    0&    0&    0&    0&    \frac{\sqrt{6}}{12}&   \frac{\sqrt{6}}{12}i &   0\\
0&    0&    0&    0&    0&    -\frac{\sqrt{6}}{12}i &   \frac{\sqrt{6}}{12}&   0\\
0&    0&    0&    -\frac{\sqrt{6}}{12}&  \frac{\sqrt{6}}{12}i &   0&    0&    0\\
0&    0&    0&    -\frac{\sqrt{6}}{12}i &   -\frac{\sqrt{6}}{12}&  0&    0&    0\\
0&    0&    0&    0&    0&    0&    0&    0\\
\end{array}\right)\nonumber
\end{eqnarray}

\begin{eqnarray}
{ C^{\bar{8}}[6]}=\left(
\begin{array}{cccccccc}
0&    0&    0&    \frac{\sqrt{6}}{12}&   -\frac{\sqrt{6}}{12}i &   0&    0&
0\\
0&    0&    0&    \frac{\sqrt{6}}{12}i &   \frac{\sqrt{6}}{12}&   0&    0&    0\\
0&    0&    0&    0&    0&    -\frac{\sqrt{6}}{12}&  \frac{\sqrt{6}}{12}i &   0\\
-\frac{\sqrt{6}}{12}&  -\frac{\sqrt{6}}{12}i &   0&    0&    0&    0&    0&    0\\
\frac{\sqrt{6}}{12}i &   -\frac{\sqrt{6}}{12}&  0&    0&    0&    0&    0&    0\\
0&    0&    \frac{\sqrt{6}}{12}&   0&    0&    0&    0&    -\frac{\sqrt{2}}{4}\\
0&    0&    -\frac{\sqrt{6}}{12}i &   0&    0&    0&    0&    \frac{\sqrt{2}}{4}i \\
0&    0&    0&    0&    0&    \frac{\sqrt{2}}{4}&   -\frac{\sqrt{2}}{4}i &   0\\
\end{array}\right)\nonumber
\end{eqnarray}

\begin{eqnarray}
{ C^{\bar{8}}[7]}=\left(
\begin{array}{cccccccc}
0&    0&    0&    \frac{\sqrt{6}}{12}&   \frac{\sqrt{6}}{12}i &   0&    0&  0\\
0&    0&    0&    -\frac{\sqrt{6}}{12}i &   \frac{\sqrt{6}}{12}&   0&    0&    0\\
0&    0&    0&    0&    0&    -\frac{\sqrt{6}}{12}&  -\frac{\sqrt{6}}{12}i &   0\\
-\frac{\sqrt{6}}{12}&  \frac{\sqrt{6}}{12}i &   0&    0&    0&    0&    0&    0\\
-\frac{\sqrt{6}}{12}i &   -\frac{\sqrt{6}}{12}&  0&    0&    0&    0&    0&    0\\
0&    0&    \frac{\sqrt{6}}{12}&   0&    0&    0&    0&    -\frac{\sqrt{2}}{4}\\
0&    0&    \frac{\sqrt{6}}{12}i &   0&    0&    0&    0&    -\frac{\sqrt{2}}{4}i \\
0&    0&    0&    0&    0&    \frac{\sqrt{2}}{4}&   \frac{\sqrt{2}}{4}i &   0\\
\end{array}\right)\nonumber
\end{eqnarray}

\begin{eqnarray}
{ C^{\bar{8}}[8]}=\left(
\begin{array}{cccccccc}
0&    -\frac{1}{2}i &   0&    0&    0&    0&    0&    0\\
\frac{1}{2}i &   0&    0&    0&    0&    0&    0&    0\\
0&    0&    0&    0&    0&    0&    0&    0\\
0&    0&    0&    0&    0&    0&    0&    0\\
0&    0&    0&    0&    0&    0&    0&    0\\
0&    0&    0&    0&    0&    0&    \frac{1}{2}i &   0\\
0&    0&    0&    0&    0&    -\frac{1}{2}i &   0&    0\\
0&    0&    0&    0&    0&    0&    0&    0\\
\end{array}\right)\nonumber
\end{eqnarray}

\begin{itemize}
  \item
In the basis
\{ $\omega^-S^+$, $hS_I$, $hS_R$, $zS_I$, $zS_R$, $(S^-S^+)^{8}$,
$\frac{(S_IS_I)^{8}}{\sqrt{2}}$, $\frac{(S_RS_R)^{8}}{\sqrt{2}}$, $(S_IS_R)^{8}$,
$(S^-S^+)^{\bar{8}}$, $(S_IS_R)^{\bar{8}}$ \},
the non-zero $A_{ij}$ is given by \\
$A_{11} = \frac{1}{2}\lambda_{1,2}^{+}$,\quad
$A_{12} =-A_{15}= \frac{i}{4}(\lambda_2-2\lambda_3)$,\quad
$A_{13} = A_{14} = \frac{1}{4}(\lambda_2+2\lambda_3)$,\quad
$A_{16} = \frac{1}{2}\sqrt[]{\frac{5}{3}}\lambda_{4,5}^{+}$,\\
$A_{17} = A_{18} = A_{37} = A_{48} = \frac{\sqrt{2}}{8}\sqrt[]{\frac{5}{3}}\lambda_{4,5}^{+}$,\quad
$A_{29} = A_{36} = A_{46} = A_{59} = \frac{1}{4}\sqrt[]{\frac{5}{3}}\lambda_{4,5}^{+}$,\\
$A_{1,11} = \frac{\sqrt{3}}{4}\lambda_{4,5}^-$,\quad
$A_{3,10} = A_{4,10} = \frac{\sqrt{3}i}{4}\lambda_{4,5}^-$,\quad
$A_{22} = A_{55} = \frac{1}{2}\lambda_{1,2}^{+}-\lambda_3$,\quad
$A_{25} = A_{34}=\lambda_3$,\\
$A_{33} = A_{44} = \frac{1}{2}\lambda_{1,2}^{+}+\lambda_3$,\quad
$A_{38} = A_{47} = \frac{1}{4\sqrt{2}}\sqrt[]{15}\lambda_{4,5}^{+}$,\quad
$A_{66} = \frac{5}{6}\lambda_{6,7}^{+}+\frac{1}{2}\lambda_{8,9}^{+}+\lambda_{10}+\frac{1}{12}\lambda_{11}$,\\
$A_{67} = A_{68} = \frac{5}{12\sqrt{2}}\lambda_{6,7,11}^{++}+\frac{1}{2\sqrt{2}}\lambda_{9,10}^{+}$,\quad
$A_{6,11} = \frac{\sqrt{5}}{4}\lambda_{6,7}^-$,\quad
$A_{77} =A_{88} = \frac{1}{4}\lambda_{6,7,11}^{++}+\frac{1}{2}\lambda_{8,9,10}^{++}$,\\
$A_{78} = \frac{7}{12}\lambda_{6,7}^++\frac{1}{2}\lambda_{10}-\frac{1}{6}\lambda_{11}$,\quad
$A_{7,10} = A_{8,10} = \frac{\sqrt{10}i}{8}\lambda_{6,7}^-$,\quad
$A_{99} = -\frac{1}{3}\lambda_{6,7}^{+}+\frac{1}{2}\lambda_{8,9}^{+}+\frac{5}{12}\lambda_{11}$,\\
$A_{10,10} = A_{13,13} = \frac{3}{2}\lambda_{6,7}^{+}+\frac{1}{2}\lambda_{8,9}^{+}-\lambda_{10}-\frac{3}{4}\lambda_{11}$,\quad
$A_{10,11} = -\frac{3i}{4}\lambda_{6,7,11}^{+-}-\frac{i}{2}\lambda_{9,10}^{-}$.
  \item
In the basis
\{ $\omega^-S_I$, $\omega^-S_R$, $hS^-$, $zS^-$, $(S_IS^-)^{8}$,
$(S_RS^-)^{8}$, $(S_IS^-)^{\bar{8}}$, $(S_RS^-)^{\bar{8}}$ \},
the non-zero $A_{ij}$ is given by \\
$A_{11} = A_{22} = A_{33} = A_{44} = \frac{1}{2}\lambda_1$,\2
$A_{13} = -A_{24} = \frac{i}{4}(\lambda_2-2\lambda_3)$,\2
$A_{14} = A_{23} = \frac{1}{4}(\lambda_2+2\lambda_3)$,\\
$A_{15} = A_{26} = A_{36} = A_{45} = \frac{1}{4}\sqrt[]{\frac{5}{3}}\lambda_{4,5}^{+}$,\quad
$A_{18} = -A_{27} = \frac{\sqrt{3}}{4}\lambda_{4,5}^-$,\quad
$A_{38} = A_{47} = \frac{i\sqrt{3}}{4}\lambda_{4,5}^-$,\\
$A_{55} = A_{66} = \frac{1}{24}\lambda_{6,7,11}^{++}+\frac{1}{2}\lambda_8+\frac{1}{4}\lambda_{9,10}^+$,\quad
$A_{56} = -\frac{3i}{8}\lambda_{6,7,11}^{+-}+\frac{i}{4}\lambda_{9,10}^-$,\quad
$A_{57} = A_{68} = \frac{i\sqrt{5}}{8}\lambda_{6,7}^-$,\\
$A_{58} = -A_{67} = \frac{\sqrt{5}}{8}\lambda_{6,7}^-$,\quad
$A_{77} = A_{88} = \frac{3}{8}\lambda_{6,7,11}^{++}-\frac{1}{4}\lambda_{9,10}^{+}+\frac{1}{2}\lambda_{8}$,\quad
$A_{78} =-
\frac{3i}{8}\lambda_{6,7,11}^{+-}-\frac{i}{4}\lambda_{9,10}^{-}$.
\end{itemize}

3. Scatterings between states in 10-dimensional representation of the SU(3) group.

In this case, the Clebsch-Gordan coefficients are given by
\begin{eqnarray}
{ C^{10}[1]}=\left(
\begin{array}{cccccccc}
0&    0&    0&    -\frac{\sqrt{2}}{4}&  -\frac{\sqrt{2}}{4}i &   0&    0&
0\\
 0&    0&    0&    -\frac{\sqrt{2}}{4}i &   \frac{\sqrt{2}}{4}&   0&    0&    0\\
 0&    0&    0&    0&    0&    0&    0&    0\\
 \frac{\sqrt{2}}{4}&   \frac{\sqrt{2}}{4}i &   0&    0&    0&    0&    0&    0\\
 \frac{\sqrt{2}}{4}i &   -\frac{\sqrt{2}}{4}&  0&    0&    0&    0&    0&    0\\
 0&    0&    0&    0&    0&    0&    0&    0\\
 0&    0&    0&    0&    0&    0&    0&    0\\
 0&    0&    0&    0&    0&    0&    0&    0\\
\end{array}\right)\nonumber
\end{eqnarray}

\begin{eqnarray}
{ C^{10}[2]}=\left(
\begin{array}{cccccccc}
0&    0&    \frac{\sqrt{6}}{12}&   0&    0&    0&    0&
\frac{\sqrt{2}}{4}\\
0&    0&    \frac{\sqrt{6}}{12}i &   0&    0&    0&    0&    \frac{\sqrt{2}}{4}i \\
-\frac{\sqrt{6}}{12}&  -\frac{\sqrt{6}}{12}i &   0&    0&    0&    0&    0&    0\\
0&    0&    0&    0&    0&    \frac{\sqrt{6}}{12}&   -\frac{\sqrt{6}}{12}i &   0\\
0&    0&    0&    0&    0&    \frac{\sqrt{6}}{12}i &   \frac{\sqrt{6}}{12}&   0\\
0&    0&    0&    -\frac{\sqrt{6}}{12}&  -\frac{\sqrt{6}}{12}i &   0&    0&    0\\
0&    0&    0&    \frac{\sqrt{6}}{12}i &   -\frac{\sqrt{6}}{12}&  0&    0&    0\\
-\frac{\sqrt{2}}{4}&  -\frac{\sqrt{2}}{4}i &   0&    0&    0&    0&    0&    0\\
\end{array}\right)\nonumber
\end{eqnarray}

\begin{eqnarray}
{ C^{10}[3]}=\left(
\begin{array}{cccccccc}
0&    0&    0&    \frac{\sqrt{6}}{12}&   -\frac{\sqrt{6}}{12}i &   0&    0&
0\\
0&    0&    0&    \frac{\sqrt{6}}{12}i &   \frac{\sqrt{6}}{12}&   0&    0&    0\\
0&    0&    0&    0&    0&    -\frac{\sqrt{6}}{12}&  \frac{\sqrt{6}}{12}i &   0\\
-\frac{\sqrt{6}}{12}&  -\frac{\sqrt{6}}{12}i &   0&    0&    0&    0&    0&    0\\
\frac{\sqrt{6}}{12}i &   -\frac{\sqrt{6}}{12}&  0&    0&    0&    0&    0&    0\\
0&    0&    \frac{\sqrt{6}}{12}&   0&    0&    0&    0&    \frac{\sqrt{2}}{4}\\
0&    0&    -\frac{\sqrt{6}}{12}i &   0&    0&    0&    0&    -\frac{\sqrt{2}}{4}i \\
0&    0&    0&    0&    0&    -\frac{\sqrt{2}}{4}&  \frac{\sqrt{2}}{4}i &   0\\
\end{array}\right)\nonumber
\end{eqnarray}

\begin{eqnarray}
{ C^{10}[4]}=\left(
\begin{array}{cccccccc}
0&    0&    0&    0&    0&    0&    0&    0\\
0&    0&    0&    0&    0&    0&    0&    0\\
0&    0&    0&    0&    0&    0&    0&    0\\
0&    0&    0&    0&    0&    -\frac{\sqrt{2}}{4}&  \frac{\sqrt{2}}{4}i &   0\\
0&    0&    0&    0&    0&    \frac{\sqrt{2}}{4}i &   \frac{\sqrt{2}}{4}&   0\\
0&    0&    0&    \frac{\sqrt{2}}{4}&   -\frac{\sqrt{2}}{4}i &   0&    0&    0\\
0&    0&    0&    -\frac{\sqrt{2}}{4}i &   -\frac{\sqrt{2}}{4}&  0&    0&    0\\
0&    0&    0&    0&    0&    0&    0&    0\\
\end{array}\right)\nonumber
\end{eqnarray}

\begin{eqnarray}
{ C^{10}[5]}=\left(
\begin{array}{cccccccc}
0&    0&    0&    0&    0&    -\frac{\sqrt{6}}{12}&  \frac{\sqrt{6}}{12}i &
0\\
0&    0&    0&    0&    0&    \frac{\sqrt{6}}{12}i &   \frac{\sqrt{6}}{12}&   0\\
0&    0&    0&    -\frac{\sqrt{6}}{12}&  \frac{\sqrt{6}}{12}i &   0&    0&    0\\
0&    0&    \frac{\sqrt{6}}{12}&   0&    0&    0&    0&    -\frac{\sqrt{2}}{4}\\
0&    0&    -\frac{\sqrt{6}}{12}i &   0&    0&    0&    0&    \frac{\sqrt{2}}{4}i \\
\frac{\sqrt{6}}{12}&   -\frac{\sqrt{6}}{12}i &   0&    0&    0&    0&    0&    0\\
-\frac{\sqrt{6}}{12}i &   -\frac{\sqrt{6}}{12}&  0&    0&    0&    0&    0&    0\\
0&    0&    0&    \frac{\sqrt{2}}{4}&   -\frac{\sqrt{2}}{4}i &   0&    0&    0\\
\end{array}\right)\nonumber
\end{eqnarray}

\begin{eqnarray}
{ C^{10}[6]}=\left(
\begin{array}{cccccccc}
0&    0&    \frac{\sqrt{6}}{12}&   0&    0&    0&    0&
-\frac{\sqrt{2}}{4}\\
0&    0&    -\frac{\sqrt{6}}{12}i &   0&    0&    0&    0&    \frac{\sqrt{2}}{4}i \\
-\frac{\sqrt{6}}{12}&  \frac{\sqrt{6}}{12}i &   0&    0&    0&    0&    0&    0\\
0&    0&    0&    0&    0&    \frac{\sqrt{6}}{12}&   \frac{\sqrt{6}}{12}i &   0\\
0&    0&    0&    0&    0&    -\frac{\sqrt{6}}{12}i &   \frac{\sqrt{6}}{12}&   0\\
0&    0&    0&    -\frac{\sqrt{6}}{12}&  \frac{\sqrt{6}}{12}i &   0&    0&    0\\
0&    0&    0&    -\frac{\sqrt{6}}{12}i &   -\frac{\sqrt{6}}{12}&  0&    0&    0\\
\frac{\sqrt{2}}{4}&   -\frac{\sqrt{2}}{4}i &   0&    0&    0&    0&    0&    0\\
\end{array}\right)\nonumber
\end{eqnarray}

\begin{eqnarray}
{ C^{10}[7]}=\left(
\begin{array}{cccccccc}
0&    0&    0&    0&    0&    \frac{\sqrt{2}}{4}&   \frac{\sqrt{2}}{4}i &
0\\
0&    0&    0&    0&    0&    -\frac{\sqrt{2}}{4}i &   \frac{\sqrt{2}}{4}&   0\\
0&    0&    0&    0&    0&    0&    0&    0\\
0&    0&    0&    0&    0&    0&    0&    0\\
0&    0&    0&    0&    0&    0&    0&    0\\
-\frac{\sqrt{2}}{4}&  \frac{\sqrt{2}}{4}i &   0&    0&    0&    0&    0&    0\\
-\frac{\sqrt{2}}{4}i &   -\frac{\sqrt{2}}{4}&  0&    0&    0&    0&    0&    0\\
0&    0&    0&    0&    0&    0&    0&    0\\
\end{array}\right)\nonumber
\end{eqnarray}

\begin{eqnarray}
{ C^{10}[8]}=\left(
\begin{array}{cccccccc}
0&    0&    0&    0&    0&    -\frac{\sqrt{6}}{12}&  -\frac{\sqrt{6}}{12}i &
0\\
0&    0&    0&    0&    0&    -\frac{\sqrt{6}}{12}i &   \frac{\sqrt{6}}{12}&   0\\
0&    0&    0&    \frac{\sqrt{6}}{6}&   \frac{\sqrt{6}}{6}i &   0&    0&    0\\
0&    0&    -\frac{\sqrt{6}}{6}&  0&    0&    0&    0&    0\\
0&    0&    -\frac{\sqrt{6}}{6}i &   0&    0&    0&    0&    0\\
\frac{\sqrt{6}}{12}&   \frac{\sqrt{6}}{12}i &   0&    0&    0&    0&    0&    0\\
\frac{\sqrt{6}}{12}i &   -\frac{\sqrt{6}}{12}&  0&    0&    0&    0&    0&    0\\
0&    0&    0&    0&    0&    0&    0&    0\\
\end{array}\right)\nonumber
\end{eqnarray}

\begin{eqnarray}
{ C^{10}[9]}=\left(
\begin{array}{cccccccc}
0&    -\frac{\sqrt{3}}{6}i &   0&    0&    0&    0&    0&    0\\
\frac{\sqrt{3}}{6}i &   0&    0&    0&    0&    0&    0&    0\\
0&    0&    0&    0&    0&    0&    0&    -\frac{1}{2}\\
0&    0&    0&    0&    \frac{\sqrt{3}}{6}i &   0&    0&    0\\
0&    0&    0&    -\frac{\sqrt{3}}{6}i &   0&    0&    0&    0\\
0&    0&    0&    0&    0&    0&    -\frac{\sqrt{3}}{6}i &   0\\
0&    0&    0&    0&    0&    \frac{\sqrt{3}}{6}i &   0&    0\\
0&    0&    \frac{1}{2}&   0&    0&    0&    0&    0\\
\end{array}\right)\nonumber
\end{eqnarray}

\begin{eqnarray}
{ C^{10}[10]}=\left(
\begin{array}{cccccccc}
0&    0&    0&    \frac{\sqrt{6}}{12}&   \frac{\sqrt{6}}{12}i &   0&    0&0\\
0&    0&    0&    -\frac{\sqrt{6}}{12}i &   \frac{\sqrt{6}}{12}&   0&    0&    0\\
0&    0&    0&    0&    0&    \frac{\sqrt{6}}{6}&   \frac{\sqrt{6}}{6}i &   0\\
-\frac{\sqrt{6}}{12}&  \frac{\sqrt{6}}{12}i &   0&    0&    0&    0&    0&    0\\
-\frac{\sqrt{6}}{12}i &   -\frac{\sqrt{6}}{12}&  0&    0&    0&    0&    0&    0\\
0&    0&    -\frac{\sqrt{6}}{6}&  0&    0&    0&    0&    0\\
0&    0&    -\frac{\sqrt{6}}{6}i &   0&    0&    0&    0&    0\\
0&    0&    0&    0&    0&    0&   0&    0\\
\end{array}\right)\nonumber
\end{eqnarray}

\begin{eqnarray}
{ C^{\bar{10}}[1]}=\left(
\begin{array}{cccccccc}
0&    0&    0&    0&    0&    \frac{\sqrt{2}}{4}&   -\frac{\sqrt{2}}{4}i & 0\\
0&    0&    0&    0&    0&    \frac{\sqrt{2}}{4}i &   \frac{\sqrt{2}}{4}&   0\\
0&    0&    0&    0&    0&    0&    0&    0\\
0&    0&    0&    0&    0&    0&    0&    0\\
0&    0&    0&    0&    0&    0&    0&    0\\
-\frac{\sqrt{2}}{4}&  -\frac{\sqrt{2}}{4}i &   0&    0&    0&    0&    0&    0\\
\frac{\sqrt{2}}{4}i &   -\frac{\sqrt{2}}{4}&  0&    0&    0&    0&    0&    0\\
0&    0&    0&    0&    0&    0&    0&    0\\
\end{array}\right)\nonumber
\end{eqnarray}

\begin{eqnarray}
{ C^{\bar{10}}[2]}=\left(
\begin{array}{cccccccc}
0&    0&    -\frac{\sqrt{6}}{12}&  0&    0&    0&    0&
\frac{\sqrt{2}}{4}\\
0&    0&    -\frac{\sqrt{6}}{12}i &   0&    0&    0&    0&    \frac{\sqrt{2}}{4}i \\
\frac{\sqrt{6}}{12}&   \frac{\sqrt{6}}{12}i &   0&    0&    0&    0&    0&    0\\
0&    0&    0&    0&    0&    -\frac{\sqrt{6}}{12}&  \frac{\sqrt{6}}{12}i &   0\\
0&    0&    0&    0&    0&    -\frac{\sqrt{6}}{12}i &   -\frac{\sqrt{6}}{12}&  0\\
0&    0&    0&    \frac{\sqrt{6}}{12}&   \frac{\sqrt{6}}{12}i &   0&    0&    0\\
0&    0&    0&    -\frac{\sqrt{6}}{12}i &   \frac{\sqrt{6}}{12}&   0&    0&    0\\
-\frac{\sqrt{2}}{4}&  -\frac{\sqrt{2}}{4}i &   0&    0&    0&    0&    0&    0\\
\end{array}\right)\nonumber
\end{eqnarray}

\begin{eqnarray}
{ C^{\bar{10}}[3]}=\left(
\begin{array}{cccccccc}
0&    0&    0&    0&    0&  -\frac{\sqrt{6}}{12}&  -\frac{\sqrt{6}}{12}i & 0\\
0&    0&    0&    0&    0&    -\frac{\sqrt{6}}{12}i &   \frac{\sqrt{6}}{12}&   0\\
0&    0&    0&    -\frac{\sqrt{6}}{12}&  -\frac{\sqrt{6}}{12}i &   0&    0&    0\\
0&    0&    \frac{\sqrt{6}}{12}&   0&    0&    0&    0&    -\frac{\sqrt{2}}{4}\\
0&    0&    \frac{\sqrt{6}}{12}i &   0&    0&    0&    0&    -\frac{\sqrt{2}}{4}i \\
\frac{\sqrt{6}}{12}&   \frac{\sqrt{6}}{12}i &   0&    0&    0&    0&    0&    0\\
\frac{\sqrt{6}}{12}i &   -\frac{\sqrt{6}}{12}&  0&    0&    0&    0&    0&    0\\
0&    0&    0&    \frac{\sqrt{2}}{4}&   \frac{\sqrt{2}}{4}i &   0&    0&    0\\
\end{array}\right)\nonumber
\end{eqnarray}

\begin{eqnarray}
{ C^{\bar{10}}[4]}=\left(
\begin{array}{cccccccc}
0&    0&    0&    0&    0&    0&    0&    0\\
0&    0&    0&    0&    0&    0&    0&    0\\
0&    0&    0&    0&    0&    0&    0&    0\\
0&    0&    0&    0&    0&    \frac{\sqrt{2}}{4}&   \frac{\sqrt{2}}{4}i &   0\\
0&    0&    0&    0&    0&    \frac{\sqrt{2}}{4}i &   -\frac{\sqrt{2}}{4}&  0\\
0&    0&    0&    -\frac{\sqrt{2}}{4}&  -\frac{\sqrt{2}}{4}i &   0&    0&    0\\
0&    0&    0&    -\frac{\sqrt{2}}{4}i &   \frac{\sqrt{2}}{4}&   0&    0&    0\\
0&    0&    0&    0&    0&    0&    0&    0\\
\end{array}\right)\nonumber
\end{eqnarray}

\begin{eqnarray}
{ C^{\bar{10}}[5]}=\left(
\begin{array}{cccccccc}
0&    0&    0&    \frac{\sqrt{6}}{12}&   \frac{\sqrt{6}}{12}i &   0&    0& 0\\
0&    0&    0&    -\frac{\sqrt{6}}{12}i &   \frac{\sqrt{6}}{12}&   0&    0&    0\\
0&    0&    0&    0&    0&    -\frac{\sqrt{6}}{12}&  -\frac{\sqrt{6}}{12}i &   0\\
-\frac{\sqrt{6}}{12}&  \frac{\sqrt{6}}{12}i &   0&    0&    0&    0&    0&    0\\
-\frac{\sqrt{6}}{12}i &   -\frac{\sqrt{6}}{12}&  0&    0&    0&    0&    0&    0\\
0&    0&    \frac{\sqrt{6}}{12}&   0&    0&    0&    0&    \frac{\sqrt{2}}{4}\\
0&    0&    \frac{\sqrt{6}}{12}i &   0&    0&    0&    0&    \frac{\sqrt{2}}{4}i \\
0&    0&    0&    0&    0&    -\frac{\sqrt{2}}{4}&  -\frac{\sqrt{2}}{4}i &   0\\
\end{array}\right)\nonumber
\end{eqnarray}

\begin{eqnarray}
{ C^{\bar{10}}[6]}=\left(
\begin{array}{cccccccc}
0&    0&    -\frac{\sqrt{6}}{12}&  0&    0&    0&    0&-\frac{\sqrt{2}}{4}\\
0&    0&    \frac{\sqrt{6}}{12}i &   0&    0&    0&    0&    \frac{\sqrt{2}}{4}i \\
\frac{\sqrt{6}}{12}&   -\frac{\sqrt{6}}{12}i &   0&    0&    0&    0&    0&    0\\
0&    0&    0&    0&    0&    -\frac{\sqrt{6}}{12}&  -\frac{\sqrt{6}}{12}i &   0\\
0&    0&    0&    0&    0&    \frac{\sqrt{6}}{12}i &   -\frac{\sqrt{6}}{12}&  0\\
0&    0&    0&    \frac{\sqrt{6}}{12}&   -\frac{\sqrt{6}}{12}i &   0&    0&    0\\
0&    0&    0&    \frac{\sqrt{6}}{12}i &   \frac{\sqrt{6}}{12}&   0&    0&    0\\
\frac{\sqrt{2}}{4}&   -\frac{\sqrt{2}}{4}i &   0&    0&    0&    0&    0&    0\\
\end{array}\right)\nonumber
\end{eqnarray}

\begin{eqnarray}
{ C^{\bar{10}}[7]}=\left(
\begin{array}{cccccccc}
0&    0&    0&    -\frac{\sqrt{2}}{4}&  \frac{\sqrt{2}}{4}i &   0&    0&
0\\
0&    0&    0&    \frac{\sqrt{2}}{4}i &   \frac{\sqrt{2}}{4}&   0&    0&    0\\
0&    0&    0&    0&    0&    0&    0&    0\\
\frac{\sqrt{2}}{4}&   -\frac{\sqrt{2}}{4}i &   0&    0&    0&    0&    0&    0\\
-\frac{\sqrt{2}}{4}i &   -\frac{\sqrt{2}}{4}&  0&    0&    0&    0&    0&    0\\
0&    0&    0&    0&    0&    0&    0&    0\\
0&    0&    0&    0&    0&    0&    0&    0\\
0&    0&    0&    0&    0&    0&    0&    0\\
\end{array}\right)\nonumber
\end{eqnarray}

\begin{eqnarray}
{ C^{\bar{10}}[8]}=\left(
\begin{array}{cccccccc}
0&    0&    0&    \frac{\sqrt{6}}{12}&   -\frac{\sqrt{6}}{12}i &   0&    0&0\\
0&    0&    0&    \frac{\sqrt{6}}{12}i &   \frac{\sqrt{6}}{12}&   0&    0&    0\\
0&    0&    0&    0&    0&    \frac{\sqrt{6}}{6}&   -\frac{\sqrt{6}}{6}i &   0\\
-\frac{\sqrt{6}}{12}&  -\frac{\sqrt{6}}{12}i &   0&    0&    0&    0&    0&    0\\
\frac{\sqrt{6}}{12}i &   -\frac{\sqrt{6}}{12}&  0&    0&    0&    0&    0&    0\\
0&    0&    -\frac{\sqrt{6}}{6}&  0&    0&    0&    0&    0\\
0&    0&    \frac{\sqrt{6}}{6}i &   0&    0&    0&    0&    0\\
0&    0&    0&    0&    0&    0&   0&    0\\
\end{array}\right)\nonumber
\end{eqnarray}

\begin{eqnarray}
{ C^{\bar{10}}[9]}=\left(
\begin{array}{cccccccc}
0&    -\frac{\sqrt{3}}{6}i &   0&    0&    0&    0&    0&    0\\
\frac{\sqrt{3}}{6}i &   0&    0&    0&    0&    0&    0&    0\\
0&    0&    0&    0&    0&    0&    0&    \frac{1}{2}\\
0&    0&    0&    0&    \frac{\sqrt{3}}{6}i &   0&    0&    0\\
0&    0&    0&    -\frac{\sqrt{3}}{6}i &   0&    0&    0&    0\\
0&    0&    0&    0&    0&    0&    -\frac{\sqrt{3}}{6}i &   0\\
0&    0&    0&    0&    0&    \frac{\sqrt{3}}{6}i &   0&    0\\
0&    0&    -\frac{1}{2}&  0&    0&    0&    0&    0\\
\end{array}\right)\nonumber
\end{eqnarray}

\begin{eqnarray}
{ C^{\bar{10}}[10]}=\left(
\begin{array}{cccccccc}
0&    0&    0&    0&    0&    -\frac{\sqrt{6}}{12}&  \frac{\sqrt{6}}{12}i &
0\\
0&    0&    0&    0&    0&    \frac{\sqrt{6}}{12}i &   \frac{\sqrt{6}}{12}&   0\\
0&    0&    0&    \frac{\sqrt{6}}{6}&   -\frac{\sqrt{6}}{6}i &   0&    0&    0\\
0&    0&    -\frac{\sqrt{6}}{6}&  0&    0&    0&    0&    0\\
0&    0&    \frac{\sqrt{6}}{6}i &   0&    0&    0&    0&    0\\
\frac{\sqrt{6}}{12}&   -\frac{\sqrt{6}}{12}i &   0&    0&    0&    0&    0&    0\\
-\frac{\sqrt{6}}{12}i &   -\frac{\sqrt{6}}{12}&  0&    0&    0&    0&    0&    0\\
0&    0&    0&    0&    0&    0&    0&    0\\
\end{array}\right)\nonumber
\end{eqnarray}

\begin{itemize}
  \item
In the basis
\{ $(S^-S^+)^{10}$, $(S_IS_R)^{10}$ \},
$A_{ij}$ is given by \\
$A_{11} = A_{22} = \frac{1}{2}\lambda_{8,9}^{+}-\lambda_{10}$,\quad
$A_{12} = -\frac{i}{2}\lambda_{9,10}^{-}$.
  \item
In the basis
\{ $(S^-S^+)^{\bar{10}}$, $(S_IS_R)^{\bar{10}}$ \},
$A_{ij}$ is given by \\
$A_{11} = A_{22} = \frac{1}{2}\lambda_{8,9}^{+}-\lambda_{10}$,\quad
$A_{12} = -\frac{i}{2}\lambda_{9,10}^{-}$.
  \item
In the basis
\{ $(S_IS^-)^{10}$,$(S_RS^-)^{10}$ \},
$A_{ij}$ is given by \\
$A_{11} = A_{22} = \frac{1}{4}\lambda_{6,7,9,10}^{---}+\frac{1}{2}\lambda_{8}$,\quad
$A_{12} = \frac{i}{4}\lambda_{6,7,9,10}^{--+}$.
  \item
In the basis
\{ $(S_IS^-)^{\bar{10}}$,$(S_RS^-)^{\bar{10}}$ \},
$A_{ij}$ is given by \\
$A_{11} = A_{22} = -\frac{1}{4}\lambda_{6,7,9,10}^{-++}+\frac{1}{2}\lambda_{8}$,\quad
$A_{12} = -\frac{i}{4}\lambda_{6,7,9,10}^{-+-}$.
\end{itemize}

4. Scatterings between states in 27-dimensional representation of the SU(3) group.

In this case, the Clebsch-Gordan coefficients are given by
\begin{eqnarray}
{ C^{27}[1]}=\left(
\begin{array}{cccccccc}
\frac{1}{2} &   \frac{1}{2}i &   0&    0&    0&    0&    0&    0\\
\frac{1}{2}i &   -\frac{1}{2}&  0&    0&    0&    0&    0&    0\\
0&    0&    0&    0&    0&    0&    0&    0\\
0&    0&    0&    0&    0&    0&    0&    0\\
0&    0&    0&    0&    0&    0&    0&    0\\
0&    0&    0&    0&    0&    0&    0&    0\\
0&    0&    0&    0&    0&    0&    0&    0\\
0&    0&    0&    0&    0&    0&    0&    0
\end{array}\right)\nonumber
\end{eqnarray}

\begin{eqnarray}
{ C^{27}[2]}=\left(
\begin{array}{cccccccc}
0&    0&    0&    -\frac{\sqrt{2}}{4}&  -\frac{\sqrt{2}}{4}i &   0&    0&
0\\
 0&    0&    0&    -\frac{\sqrt{2}}{4}i &   \frac{\sqrt{2}}{4}&   0&    0&    0\\
 0&    0&    0&    0&    0&    0&    0&    0\\
 -\frac{\sqrt{2}}{4}&  -\frac{\sqrt{2}}{4}i &   0&    0&    0&    0&    0&    0\\
 -\frac{\sqrt{2}}{4}i &   \frac{\sqrt{2}}{4}&   0&    0&    0&    0&    0&    0\\
 0&    0&    0&    0&    0&    0&    0&    0\\
 0&    0&    0&    0&    0&    0&    0&    0\\
 0&    0&    0&    0&    0&    0&    0&    0\\
\end{array}\right)\nonumber
\end{eqnarray}

\begin{eqnarray}
{ C^{27}[3]}=\left(
\begin{array}{cccccccc}
0&    0&    0&    0&    0&    0&    0&    0\\
 0&    0&    0&    0&    0&    0&    0&    0\\
 0&    0&    0&    0&    0&    0&    0&    0\\
 0&    0&    0&    \frac{1}{2}&   \frac{1}{2}i &   0&    0&    0\\
 0&    0&    0&    \frac{1}{2}i &   -\frac{1}{2}&  0&    0&    0\\
 0&    0&    0&    0&    0&    0&    0&    0\\
 0&    0&    0&    0&    0&    0&    0&    0\\
 0&    0&    0&    0&    0&    0&    0&    0\\
\end{array}\right)\nonumber
\end{eqnarray}

\begin{eqnarray}
{ C^{27}[4]}=\left(
\begin{array}{cccccccc}
0&    0&    0&    0&    0&    0&    0&    0\\
0&    0&    0&    0&    0&    0&    0&    0\\
0&    0&    0&    -\frac{1}{4}&  -\frac{1}{4}i &   0&    0&    0\\
0&    0&    -\frac{1}{4}&  0&    0&    0&    0&    -\frac{\sqrt{3}}{4}\\
0&    0&    -\frac{1}{4}i &   0&    0&    0&    0&    -\frac{\sqrt{3}}{4} i \\
0&    0&    0&    0&    0&    0&    0&    0\\
0&    0&    0&    0&    0&    0&    0&    0\\
0&    0&    0&    -\frac{\sqrt{3}}{4}&  -\frac{\sqrt{3}}{4}i &   0&    0&    0\\
\end{array}\right)\nonumber
\end{eqnarray}

\begin{eqnarray}
{ C^{27}[5]}=\left(
\begin{array}{cccccccc}
0&    0&    0&    0&    0&    0&    0&    0\\
 0&    0&    0&    0&    0&    0&    0&    0\\
 0&    0&    \frac{\sqrt{6}}{12}&   0&    0&    0&    0&    \frac{\sqrt{2}}{4}\\
 0&    0&    0&    -\frac{\sqrt{6}}{6}&  0&    0&    0&    0\\
 0&    0&    0&    0&    -\frac{\sqrt{6}}{6}&  0&    0&    0\\
 0&    0&    0&    0&    0&    0&    0&    0\\
 0&    0&    0&    0&    0&    0&    0&    0\\
 0&    0&    \frac{\sqrt{2}}{4}&   0&    0&    0&    0&    \frac{\sqrt{6}}{4}\\
\end{array}\right)\nonumber
\end{eqnarray}

\begin{eqnarray}
{ C^{27}[6]}=\left(
\begin{array}{cccccccc}
0&    0&    0&    0&    0&    0&    0&    0\\
0&    0&    0&    0&    0&    0&    0&    0\\
0&    0&    0&    \frac{1}{4} &  -\frac{1}{4}i &   0&    0&    0\\
0&    0&    \frac{1}{4}   &  0&    0&    0&    0&    \frac{\sqrt{3}}{4}\\
0&    0&    -\frac{1}{4}i &   0&    0&    0&    0&    -\frac{\sqrt{3}}{4}i \\
0&    0&    0&    0&    0&    0&    0&    0\\
0&    0&    0&    0&    0&    0&    0&    0\\
0&    0&    0&    \frac{\sqrt{3}}{4}&   -\frac{\sqrt{3}}{4}i &   0&    0&    0\\
\end{array}\right)\nonumber
\end{eqnarray}

\begin{eqnarray}
{ C^{27}[7]}=\left(
\begin{array}{cccccccc}
0&    0&    0&    0&    0&    0&    0&    0\\
 0&    0&    0&    0&    0&    0&    0&    0\\
 0&    0&    0&    0&    0&    0&    0&    0\\
 0&    0&    0&    \frac{1}{2}&   -\frac{1}{2}i &   0&    0&    0\\
 0&    0&    0&    -\frac{1}{2}i &   -\frac{1}{2}&  0&    0&    0\\
 0&    0&    0&    0&    0&    0&    0&    0\\
 0&    0&    0&    0&    0&    0&    0&    0\\
 0&    0&    0&    0&    0&    0&    0&    0\\
\end{array}\right)\nonumber
\end{eqnarray}

\begin{eqnarray}
{ C^{27}[8]}=\left(
\begin{array}{cccccccc}
0&    0&    0&    \frac{\sqrt{2}}{4}&   -\frac{\sqrt{2}}{4}i &   0&    0&
0\\
 0&    0&    0&    -\frac{\sqrt{2}}{4}i &   -\frac{\sqrt{2}}{4}&  0&    0&    0\\
 0&    0&    0&    0&    0&    0&    0&    0\\
 \frac{\sqrt{2}}{4}&   -\frac{\sqrt{2}}{4}i &   0&    0&    0&    0&    0&    0\\
 -\frac{\sqrt{2}}{4}i &   -\frac{\sqrt{2}}{4}&  0&    0&    0&    0&    0&    0\\
 0&    0&    0&    0&    0&    0&    0&    0\\
 0&    0&    0&    0&    0&    0&    0&    0\\
 0&    0&    0&    0&    0&    0&    0&    0\\
\end{array}\right)\nonumber
\end{eqnarray}

\begin{eqnarray}
{ C^{27}[9]}=\left(
\begin{array}{cccccccc}
\frac{1}{2}&   -\frac{1}{2}i &   0&    0&    0&    0&    0&    0\\
 -\frac{1}{2}i &   -\frac{1}{2}&  0&    0&    0&    0&    0&    0\\
 0&    0&    0&    0&    0&    0&    0&    0\\
 0&    0&    0&    0&    0&    0&    0&    0\\
 0&    0&    0&    0&    0&    0&    0&    0\\
 0&    0&    0&    0&    0&    0&    0&    0\\
 0&    0&    0&    0&    0&    0&    0&    0\\
 0&    0&    0&    0&    0&    0&    0&    0\\
\end{array}\right)\nonumber
\end{eqnarray}

\begin{eqnarray}
{ C^{27}[10]}=\left(
\begin{array}{cccccccc}
0&    0&    0&    0&    0&    \frac{\sqrt{2}}{4}&   -\frac{\sqrt{2}}{4}i &
0\\
 0&    0&    0&    0&    0&    \frac{\sqrt{2}}{4}i &   \frac{\sqrt{2}}{4}&   0\\
 0&    0&    0&    0&    0&    0&    0&    0\\
 0&    0&    0&    0&    0&    0&    0&    0\\
 0&    0&    0&    0&    0&    0&    0&    0\\
 \frac{\sqrt{2}}{4}&   \frac{\sqrt{2}}{4}i &   0&    0&    0&    0&    0&    0\\
 -\frac{\sqrt{2}}{4}i &   \frac{\sqrt{2}}{4}&   0&    0&    0&    0&    0&    0\\
 0&    0&    0&    0&    0&    0&    0&    0\\
\end{array}\right)\nonumber
\end{eqnarray}

\begin{eqnarray}
{ C^{27}[11]}=\left(
\begin{array}{cccccccc}
0&    0&    0&    0&    0&    0&    0&    0\\
 0&    0&    0&    0&    0&    0&    0&    0\\
 0&    0&    0&    0&    0&    0&    0&    0\\
 0&    0&    0&    0&    0&    0&    0&    0\\
 0&    0&    0&    0&    0&    0&    0&    0\\
 0&    0&    0&    0&    0&    \frac{1}{2}&   -\frac{1}{2}i &   0\\
 0&    0&    0&    0&    0&    -\frac{1}{2}i &   -\frac{1}{2}&  0\\
 0&    0&    0&    0&    0&    0&    0&    0\\
\end{array}\right)\nonumber
\end{eqnarray}

\begin{eqnarray}
{ C^{27}[12]}=\left(
\begin{array}{cccccccc}
 0&    0&    \frac{\sqrt{6}}{12}&   0&    0&    0&    0&  \frac{\sqrt{2}}{4}\\
 0&    0&    \frac{\sqrt{6}}{12}i &   0&    0&    0&    0&    \frac{\sqrt{2}}{4}i \\
 \frac{\sqrt{6}}{12}&   \frac{\sqrt{6}}{12}i &   0&    0&    0&    0&    0&    0\\
 0&    0&    0&    0&    0&    -\frac{\sqrt{6}}{12}&  \frac{\sqrt{6}}{12}i &   0\\
 0&    0&    0&    0&    0&    -\frac{\sqrt{6}}{12}i &   -\frac{\sqrt{6}}{12}&  0\\
 0&    0&    0&    -\frac{\sqrt{6}}{12}&  -\frac{\sqrt{6}}{12}i &   0&    0&    0\\
 0&    0&    0&    \frac{\sqrt{6}}{12}i &   -\frac{\sqrt{6}}{12}&  0&    0&    0\\
 \frac{\sqrt{2}}{4}&   \frac{\sqrt{2}}{4}i &   0&    0&    0&    0&    0&    0\\
\end{array}\right)\nonumber
\end{eqnarray}

\begin{eqnarray}
{ C^{27}[13]}=\left(
\begin{array}{cccccccc}
0&    0&    0&    \frac{\sqrt{6}}{12}&   -\frac{\sqrt{6}}{12}i &   0&    0&
0\\
 0&    0&    0&    \frac{\sqrt{6}}{12}i &   \frac{\sqrt{6}}{12}&   0&    0&    0\\
 0&    0&    0&    0&    0&    \frac{\sqrt{6}}{12}&   -\frac{\sqrt{6}}{12}i &   0\\
 \frac{\sqrt{6}}{12}&   \frac{\sqrt{6}}{12}i &   0&    0&    0&    0&    0&    0\\
 -\frac{\sqrt{6}}{12}i &   \frac{\sqrt{6}}{12}&   0&    0&    0&    0&    0&    0\\
 0&    0&    \frac{\sqrt{6}}{12}&   0&    0&    0&    0&    \frac{\sqrt{2}}{4}\\
 0&    0&    -\frac{\sqrt{6}}{12}i &   0&    0&    0&    0&    -\frac{\sqrt{2}}{4}i \\
 0&    0&    0&    0&    0&    \frac{\sqrt{2}}{4}&   -\frac{\sqrt{2}}{4}i &   0\\
\end{array}\right)\nonumber
\end{eqnarray}

\begin{eqnarray}
{ C^{27}[14]}=\left(
\begin{array}{cccccccc}
0&    0&    0&    0&    0&    0&    0&    0\\
 0&    0&    0&    0&    0&    0&    0&    0\\
 0&    0&    0&    0&    0&    0&    0&    0\\
 0&    0&    0&    0&    0&    \frac{\sqrt{2}}{4}&   -\frac{\sqrt{2}}{4}i &   0\\
 0&    0&    0&    0&    0&    -\frac{\sqrt{2}}{4}i &   -\frac{\sqrt{2}}{4}&  0\\
 0&    0&    0&    \frac{\sqrt{2}}{4}&   -\frac{\sqrt{2}}{4}i &   0&    0&    0\\
 0&    0&    0&    -\frac{\sqrt{2}}{4}i &   -\frac{\sqrt{2}}{4}&  0&    0&    0\\
 0&    0&    0&    0&    0&    0&    0&    0\\
\end{array}\right)\nonumber
\end{eqnarray}

\begin{eqnarray}
{ C^{27}[15]}=\left(
\begin{array}{cccccccc}
0&    0&    0&    0&    0&    0&    0&    0\\
 0&    0&    0&    0&    0&    0&    0&    0\\
 0&    0&    0&    0&    0&    0&    0&    0\\
 0&    0&    0&    0&    0&    \frac{\sqrt{2}}{4}&   \frac{\sqrt{2}}{4}i &   0\\
 0&    0&    0&    0&    0&    \frac{\sqrt{2}}{4}i &   -\frac{\sqrt{2}}{4}&  0\\
 0&    0&    0&    \frac{\sqrt{2}}{4}&   \frac{\sqrt{2}}{4}i &   0&    0&    0\\
 0&    0&    0&    \frac{\sqrt{2}}{4}i &   -\frac{\sqrt{2}}{4}&  0&    0&    0\\
 0&    0&    0&    0&    0&    0&    0&    0\\
\end{array}\right)\nonumber
\end{eqnarray}

\begin{eqnarray}
{ C^{27}[16]}=\left(
\begin{array}{cccccccc}
0&    0&    0&    -\frac{\sqrt{6}}{12}&  -\frac{\sqrt{6}}{12}i &   0&    0&
0\\
 0&    0&    0&    \frac{\sqrt{6}}{12}i &   -\frac{\sqrt{6}}{12}&  0&    0&    0\\
 0&    0&    0&    0&    0&    -\frac{\sqrt{6}}{12}&  -\frac{\sqrt{6}}{12}i &   0\\
 -\frac{\sqrt{6}}{12}&  \frac{\sqrt{6}}{12}i &   0&    0&    0&    0&    0&    0\\
 -\frac{\sqrt{6}}{12}i &   -\frac{\sqrt{6}}{12}&  0&    0&    0&    0&    0&    0\\
 0&    0&    -\frac{\sqrt{6}}{12}&  0&    0&    0&    0&    -\frac{\sqrt{2}}{4}\\
 0&    0&    -\frac{\sqrt{6}}{12}i &   0&    0&    0&    0&    -\frac{\sqrt{2}}{4}i \\
 0&    0&    0&    0&    0&    -\frac{\sqrt{2}}{4}&  -\frac{\sqrt{2}}{4}i &   0\\
\end{array}\right)\nonumber
\end{eqnarray}

\begin{eqnarray}
{ C^{27}[17]}=\left(
\begin{array}{cccccccc}
0&    0&    \frac{\sqrt{6}}{12}&   0&    0&    0&    0&
\frac{\sqrt{2}}{4}\\
 0&    0&    -\frac{\sqrt{6}}{12}i &   0&    0&    0&    0&    -\frac{\sqrt{2}}{4}i \\
 \frac{\sqrt{6}}{12}&   -\frac{\sqrt{6}}{12}i &   0&    0&    0&    0&    0&    0\\
 0&    0&    0&    0&    0&    -\frac{\sqrt{6}}{12}&  -\frac{\sqrt{6}}{12}i &   0\\
 0&    0&    0&    0&    0&    \frac{\sqrt{6}}{12}i &   -\frac{\sqrt{6}}{12}&  0\\
 0&    0&    0&    -\frac{\sqrt{6}}{12}&  \frac{\sqrt{6}}{12}i &   0&    0&    0\\
 0&    0&    0&    -\frac{\sqrt{6}}{12}i &   -\frac{\sqrt{6}}{12}&  0&    0&    0\\
 \frac{\sqrt{2}}{4}&   -\frac{\sqrt{2}}{4}i &   0&    0&    0&    0&    0&    0\\
\end{array}\right)\nonumber
\end{eqnarray}

\begin{eqnarray}
{ C^{27}[18]}=\left(
\begin{array}{cccccccc}
0&    0&    0&    0&    0&    0&    0&    0\\
 0&    0&    0&    0&    0&    0&    0&    0\\
 0&    0&    0&    0&    0&    0&    0&    0\\
 0&    0&    0&    0&    0&    0&    0&    0\\
 0&    0&    0&    0&    0&    0&    0&    0\\
 0&    0&    0&    0&    0&    \frac{1}{2}&   \frac{1}{2}i &   0\\
 0&    0&    0&    0&    0&    \frac{1}{2}i &   -\frac{1}{2}&  0\\
 0&    0&    0&    0&    0&    0&    0&    0\\
\end{array}\right)\nonumber
\end{eqnarray}

\begin{eqnarray}
{ C^{27}[19]}=\left(
\begin{array}{cccccccc}
0&    0&    0&    0&    0&    -\frac{\sqrt{2}}{4}&  -\frac{\sqrt{2}}{4}i &
0\\
 0&    0&    0&    0&    0&    \frac{\sqrt{2}}{4}i &   -\frac{\sqrt{2}}{4}&  0\\
 0&    0&    0&    0&    0&    0&    0&    0\\
 0&    0&    0&    0&    0&    0&    0&    0\\
 0&    0&    0&    0&    0&    0&    0&    0\\
 -\frac{\sqrt{2}}{4}&  \frac{\sqrt{2}}{4}i &   0&    0&    0&    0&    0&    0\\
 -\frac{\sqrt{2}}{4}i &   -\frac{\sqrt{2}}{4}&  0&    0&    0&    0&    0&    0\\
 0&    0&    0&    0&    0&    0&    0&    0\\
\end{array}\right)\nonumber
\end{eqnarray}

\begin{eqnarray}
{ C^{27}[20]}=\left(
\begin{array}{cccccccc}
0&    0&    -\frac{\sqrt{30}}{12}&  0&    0&    0&    0&
\frac{\sqrt{10}}{20}\\
0&    0&    -\frac{\sqrt{30}}{12}i &   0&    0&    0&    0&    \frac{\sqrt{10}}{20}i \\
-\frac{\sqrt{30}}{12}&  -\frac{\sqrt{30}}{12}i &   0&    0&    0&    0&    0&    0\\
0&    0&    0&    0&    0&    -\frac{\sqrt{30}}{60}& \frac{\sqrt{30}}{60}i &  0\\
0&    0&    0&    0&    0&    -\frac{\sqrt{30}}{60}i &  -\frac{\sqrt{30}}{60}& 0\\
0&    0&    0&    -\frac{\sqrt{30}}{60}& -\frac{\sqrt{30}}{60}i &  0&    0&    0\\
0&    0&    0&    \frac{\sqrt{30}}{60}i &  -\frac{\sqrt{30}}{60}& 0&    0&    0\\
\frac{\sqrt{10}}{20}&   \frac{\sqrt{10}}{20}i &   0&    0&    0&    0&    0&    0\\
\end{array}\right)\nonumber
\end{eqnarray}

\begin{eqnarray}
{ C^{27}[21]}=\left(
\begin{array}{cccccccc}
0&    0&    0&    -\frac{\sqrt{30}}{60}& \frac{\sqrt{30}}{60}i &  0&    0&
0\\
 0&    0&    0&    -\frac{\sqrt{30}}{60}i &  -\frac{\sqrt{30}}{60}& 0&    0&    0\\
 0&    0&    0&    0&    0&    -\frac{\sqrt{30}}{15}&  \frac{\sqrt{30}}{15}i &   0\\
 -\frac{\sqrt{30}}{60}& -\frac{\sqrt{30}}{60}i &  0&    0&    0&    0&    0&    0\\
 \frac{\sqrt{30}}{60}i &  -\frac{\sqrt{30}}{60}& 0&    0&    0&    0&    0&    0\\
 0&    0&    -\frac{\sqrt{30}}{15}&  0&    0&    0&    0&    \frac{\sqrt{10}}{10}\\
 0&    0&    \frac{\sqrt{30}}{15}i &   0&    0&    0&    0&    -\frac{\sqrt{10}}{10}i \\
 0&    0&    0&    0&    0&    \frac{\sqrt{10}}{10}&   -\frac{\sqrt{10}}{10}i &   0\\
\end{array}\right)\nonumber
\end{eqnarray}

\begin{eqnarray}
{ C^{27}[22]}=\left(
\begin{array}{cccccccc}
0&    0&    0&    0&    0&    -\frac{\sqrt{5}}{10}&  -\frac{\sqrt{5}}{10}i &
0\\
 0&    0&    0&    0&    0&    -\frac{\sqrt{5}}{10}i &   \frac{\sqrt{5}}{10}&   0\\
 0&    0&    0&    \frac{3\sqrt{5}}{20}&   \frac{3\sqrt{5}}{20}i &   0&    0&    0\\
 0&    0&    \frac{3\sqrt{5}}{20}&   0&    0&    0&    0&    -\frac{\sqrt{15}}{20}\\
 0&    0&    \frac{3\sqrt{5}}{20}i &   0&    0&    0&    0&    -\frac{\sqrt{15}}{20}i \\
 -\frac{\sqrt{5}}{10}&  -\frac{\sqrt{5}}{10}i &   0&    0&    0&    0&    0&    0\\
 -\frac{\sqrt{5}}{10}i &   \frac{\sqrt{5}}{10}&   0&    0&    0&    0&    0&    0\\
 0&    0&    0&    -\frac{\sqrt{15}}{20}&  -\frac{\sqrt{15}}{20}i &   0&    0&    0\\
\end{array}\right)\nonumber
\end{eqnarray}

\begin{eqnarray}
{ C^{27}[23]}=\left(
\begin{array}{cccccccc}
\frac{\sqrt{10}}{10}&   0&    0&    0&    0&    0&    0&    0\\
 0&    \frac{\sqrt{10}}{10}&   0&    0&    0&    0&    0&    0\\
 0&    0&    -\frac{3\sqrt{10}}{20}&  0&    0&    0&    0&    -\frac{\sqrt{30}}{20}\\
 0&    0&    0&    0&    0&    0&    0&    0\\
 0&    0&    0&    0&    0&    0&    0&    0\\
 0&    0&    0&    0&    0&    -\frac{\sqrt{10}}{10}&  0&    0\\
 0&    0&    0&    0&    0&    0&    -\frac{\sqrt{10}}{10}&  0\\
 0&    0&    -\frac{\sqrt{30}}{20}&  0&    0&    0&    0&    \frac{3\sqrt{10}}{20}\\
\end{array}\right)\nonumber
\end{eqnarray}

\begin{eqnarray}
{ C^{27}[24]}=\left(
\begin{array}{cccccccc}
0&    0&    0&    0&    0&    \frac{\sqrt{5}}{10}&   -\frac{\sqrt{5}}{10}i &
0\\
 0&    0&    0&    0&    0&    -\frac{\sqrt{5}}{10}i &   -\frac{\sqrt{5}}{10}&  0\\
 0&    0&    0&    -\frac{3\sqrt{5}}{20}&  \frac{3\sqrt{5}}{20}i &   0&    0&    0\\
 0&    0&    -\frac{3\sqrt{5}}{20}&  0&    0&    0&    0&    \frac{\sqrt{15}}{20}\\
 0&    0&    \frac{3\sqrt{5}}{20}i &   0&    0&    0&    0&    -\frac{\sqrt{15}}{20}i \\
 \frac{\sqrt{5}}{10}&   -\frac{\sqrt{5}}{10}i &   0&    0&    0&    0&    0&    0\\
 -\frac{\sqrt{5}}{10}i &   -\frac{\sqrt{5}}{10}&  0&    0&    0&    0&    0&    0\\
 0&    0&    0&    \frac{\sqrt{15}}{20}&   -\frac{\sqrt{15}}{20}i &   0&    0&    0\\
\end{array}\right)\nonumber
\end{eqnarray}

\begin{eqnarray}
{ C^{27}[25]}=\left(
\begin{array}{cccccccc}
-\frac{\sqrt{30}}{20}&  0&    0&    0&    0&    0&    0&    0\\
 0&    -\frac{\sqrt{30}}{20}&  0&    0&    0&    0&    0&    0\\
 0&    0&    \frac{7\sqrt{30}}{60}&   0&    0&    0&    0&    -\frac{\sqrt{10}}{10}\\
 0&    0&    0&    \frac{\sqrt{30}}{60}&  0&    0&    0&    0\\
 0&    0&    0&    0&    \frac{\sqrt{30}}{60}&  0&    0&    0\\
 0&    0&    0&    0&    0&    -\frac{\sqrt{30}}{20}&  0&    0\\
 0&    0&    0&    0&    0&    0&    -\frac{\sqrt{30}}{20}&  0\\
 0&    0&    -\frac{\sqrt{10}}{10}&  0&    0&    0&    0&    \frac{\sqrt{30}}{20}\\
\end{array}\right)\nonumber
\end{eqnarray}

\begin{eqnarray}
{ C^{27}[26]}=\left(
\begin{array}{cccccccc}
0&    0&    0&    \frac{\sqrt{30}}{60}&  \frac{\sqrt{30}}{60}i &  0&    0&
0\\
 0&    0&    0&    -\frac{\sqrt{30}}{60}i &  \frac{\sqrt{30}}{60}&  0&    0&    0\\
 0&    0&    0&    0&    0&    \frac{\sqrt{30}}{15}&   \frac{\sqrt{30}}{15}i &   0\\
 \frac{\sqrt{30}}{60}&  -\frac{\sqrt{30}}{60}i &  0&    0&    0&    0&    0&    0\\
 \frac{\sqrt{30}}{60}i &  \frac{\sqrt{30}}{60}&  0&    0&    0&    0&    0&    0\\
 0&    0&    \frac{\sqrt{30}}{15}&   0&    0&    0&    0&    -\frac{\sqrt{10}}{10}\\
 0&    0&    \frac{\sqrt{30}}{15}i &   0&    0&    0&    0&    -\frac{\sqrt{10}}{10}i \\
 0&    0&    0&    0&    0&    -\frac{\sqrt{10}}{10}&  -\frac{\sqrt{10}}{10}i &   0\\
\end{array}\right)\nonumber
\end{eqnarray}

\begin{eqnarray}
{ C^{27}[27]}=\left(
\begin{array}{cccccccc}
0&    0&    -\frac{\sqrt{30}}{12}&  0&    0&    0&    0&
\frac{\sqrt{10}}{20}\\
 0&    0&    \frac{\sqrt{30}}{12}i &   0&    0&    0&    0&    -\frac{\sqrt{10}}{20}i \\
 -\frac{\sqrt{30}}{12}&  \frac{\sqrt{30}}{12}i &   0&    0&    0&    0&    0&    0\\
 0&    0&    0&    0&    0&    -\frac{\sqrt{30}}{60}& -\frac{\sqrt{30}}{60}i &  0\\
 0&    0&    0&    0&    0&    \frac{\sqrt{30}}{60}i &  -\frac{\sqrt{30}}{60}& 0\\
 0&    0&    0&    -\frac{\sqrt{30}}{60}& \frac{\sqrt{30}}{60}i &  0&    0&    0\\
 0&    0&    0&    -\frac{\sqrt{30}}{60}i &  -\frac{\sqrt{30}}{60}& 0&    0&    0\\
 \frac{\sqrt{10}}{20}&   -\frac{\sqrt{10}}{20}i &   0&    0&    0&    0&    0&    0\\
\end{array}\right)\nonumber
\end{eqnarray}

\begin{itemize}
  \item
In the basis
\{ $(S^-S^+)^{27}$, $\frac{(S_IS_I)^{27}}{\sqrt{2}}$,
$\frac{(S_RS_R)^{27}}{\sqrt{2}}$ \},
$A_{ij}$ is given by \\
$A_{11} = \frac{1}{2}\lambda_{8,9,11}^{++}+\lambda_{10}$,\quad
$A_{12} = A_{13} = \frac{\sqrt{2}}{4}\lambda_{9,10}^{+}$,\quad
$A_{22} = A_{33} = \frac{1}{4}\lambda_{6,7,11}^{++}+\frac{1}{2}\lambda_{8,9,10}^{++}$,\\
$A_{23} = -\frac{1}{4}\lambda_{6,7,11}^{+-}+\frac{1}{2}\lambda_{10}$.
  \item
In the basis \{ $(S_IS_R)^{27}$ \}, the S-partial wave amplitude for the scattering $(S_IS_R)^{27} \to (S_IS_R)^{27}$ is \quad
$A_{11} = \frac{1}{2}\lambda_{6,7,8,9}^{+++}$.
  \item
In the basis \{ $(S_IS^-)^{27}$,$(S_RS^-)^{27}$ \}, $A_{ij}$ is given by\\
$A_{11} = A_{22} = \frac{1}{4}\lambda_{6,7,9,10,11}^{++++}
         +\frac{1}{2}\lambda_{8}$,\quad
$A_{12} = \frac{i}{4}\lambda_{6,7,9,10,11}^{++--}$.
\end{itemize}

\end{document}